\documentclass[fleqn,11pt,twoside]{article}

\usepackage{amsthm,amsthm,amssymb,color,xcolor,epsfig,graphics,subfigure}

\usepackage{amsmath, graphicx, latexsym, lscape, cite }

\makeatletter
\newcommand{\copyrightnote}[2]{{\renewcommand{\thefootnote}{}
 \footnotetext{\small\it
\begin{flushleft}
 \copyright \ #1   #2  
\end{flushleft}}}}
\def\d{\mbox{\rm d}}

\newcommand{\Name}[1]{\begin{flushleft}
                       \LARGE \bf #1
                       \end{flushleft}\vspace{-3mm}}

\newcommand{\Author}[1]{\begin{flushleft}
                       \it #1 \end{flushleft}}

\newcommand{\Address}[1]{\begin{flushleft}
                       \it #1 \end{flushleft}}

\newcommand{\Date}[1]{\begin{flushleft}
                      \small  \it #1 \end{flushleft}}

\newcommand{\evenhead}{G. Gaeta}
\newcommand{\oddhead}{Symmetry of the isotropic Ornstein-Uhlenbeck process in a force field}

\renewcommand{\@evenhead}{
\hspace*{-3pt}\raisebox{-15pt}[\headheight][0pt]{\vbox{\hbox to \textwidth
{\thepage \hfil \evenhead}\vskip4pt \hrule}}}
\renewcommand{\@oddhead}{
\hspace*{-3pt}\raisebox{-15pt}[\headheight][0pt]{\vbox{\hbox to \textwidth
{\oddhead \hfil \thepage}\vskip4pt\hrule}}}
\renewcommand{\@evenfoot}{}
\renewcommand{\@oddfoot}{}

\long\def\@makecaption#1#2{%
  \vskip\abovecaptionskip
  \sbox\@tempboxa{\small \textbf{#1.}\ \ #2}%
  \ifdim \wd\@tempboxa >\hsize
    {\small \textbf{#1.}\ \ #2}\par
  \else
    \global \@minipagefalse
    \hb@xt@\hsize{\hfil\box\@tempboxa\hfil}%
  \fi
  \vskip\belowcaptionskip}

\newcommand{\JNMPnumberwithin}[3][\arabic]{%
  \@ifundefined{c@#2}{\@nocounterr{#2}}{%
    \@ifundefined{c@#3}{\@nocnterr{#3}}{%
      \@addtoreset{#2}{#3}%
      \@xp\xdef\csname the#2\endcsname{%
        \@xp\@nx\csname the#3\endcsname .\@nx#1{#2}}}}%
}

\newcommand{\resetfootnoterule} {
  \renewcommand\footnoterule{%
  \kern-3\p@
  \hrule\@width.4\columnwidth
  \kern2.6\p@}
}

\renewcommand{\footnoterule}{}

\makeatother

\theoremstyle{definition}

\begin{document}

\renewcommand{\evenhead}{ {\LARGE\textcolor{blue!10!black!40!green}{{\sf \ \ \ ]ocnmp[}}}\strut\hfill G. Gaeta}
\renewcommand{\oddhead}{ {\LARGE\textcolor{blue!10!black!40!green}{{\sf ]ocnmp[}}}\ \ \ \ \  Symmetry of the OU process}

%%%% Matter for the first page 
\thispagestyle{empty}
\newcommand{\FistPageHead}[3]{
\begin{flushleft}
\raisebox{8mm}[0pt][0pt]
{\footnotesize \sf
\parbox{150mm}{{Open Communications in Nonlinear Mathematical Physics}\ \  \ {\LARGE\textcolor{blue!10!black!40!green}{]ocnmp[}}
\ \ Vol.1 (2021) pp
#2\hfill {\sc #3}}}\vspace{-13mm}
\end{flushleft}}

\def\a{\alpha}
\def\b{\beta}
\def\ga{\gamma}
\def\de{\delta}
\def\De{\Delta}
\def\la{\lambda}
\def\La{\Lambda}
\def\s{\sigma}
\def\om{\omega}
\def\vphi{\varphi}
\def\eps{\varepsilon}

\def\xb{{\bf x}}
\def\yb{{\bf y}}
\def\ub{{\bf u}}
\def\vb{{\bf v}}
\def\wb{{\bf w}}
\def\zb{{\bf z}}
\def\fb{{\bf f}}
\def\pb{{\bf p}}
\def\qb{{\bf q}}
\def\yb{{\bf y}}
\def\Fb{{\bf F}}
\def\Rb{{\bf R}}

\def\xd{\dot{x}}
\def\xdd{\ddot{x}}

\def\zetab{\zeta}

\def\L{\mathcal{L}}
\def\G{\mathcal{G}}

\def\grad{\nabla}

\def\pa{\partial}
\def\d{{\mathrm d}}
\def\T{\mathrm{T}}
\def\E{\mathcal{E}}

\def\X{\mathcal{X}}
\def\Y{\mathcal{Y}}
\def\S{\mathcal{S}}

\def\^#1{\widehat{#1}}
\def\wh#1{\widehat{#1}}
\def\wt#1{\widetilde{#1}}

\def\({\left(}
\def\){\right)}
\def\[{\left[}
\def\]{\right]}

\def\rosso{\color{red}}
\def\ro{\color{red}}
\def\blu{\color{blue}}
\def\verde{\color{green}}
\def\arancio{\color{orange}}
\def\viola{\color{violet}}

\def\gcite#1{{\blu \cite{#1}}}

\def\beq{\begin{equation}}
\def\eeq{\end{equation}}

\def\EOR{\hfill $\odot$}

\def\beql#1{\begin{equation} \label{#1}}

\def\eqref#1{(\ref{#1})}

\def\symmref{AVL,CGbook,KrV,Olver1,Olver2,Stephani,Win}
\def\sderef{Arnold,Evans,Fre,Ikeda,Kampen,Oksendal,Stroock}
\def\dsref{GH,Gle,IA,Ver}
\def\stochsymmref{AlbFei,Mis1,Mis2,Mis3,GRQ1,GRQ2,Unal,Koz1,Koz2,Koz3,GS17,GGPR,GL1,GL2,Koz18a,Koz18b,KozB,GLS,GW18,Glogistic,GSclass, DVMU0,DVMU1,DVMU2,DVMU3,DVMU4}

\FistPageHead{1}{\pageref{firstpage}--\pageref{lastpage}}{ \ \ Article}

\strut\hfill

\strut\hfill

\copyrightnote{The author(s). Distributed under a Creative Commons Attribution 4.0 International License}

\Name{Symmetry of the isotropic Ornstein-Uhlenbeck process in a force field}

\Author{Giuseppe Gaeta$^{\,1,2,3}$}

\Address{$^{1}$ {\it Dipartimento di Matematica, Universit\`a degli Studi di Milano,  v. Saldini 50, I-20133 Milano (Italy)}\\[2mm]
$^{2}$ {\it SMRI, Santa Marinella (Italy)} \\[2mm]
$^{3}$ {\it GNFM-INdAM (Italy)} \\[2mm] {\tt giuseppe.gaeta@unimi.it}}

\Date{Received June 2, 2021; Accepted November 14, 2021}

\setcounter{equation}{0}

\begin{abstract}
\noindent
We classify simple symmetries for an Ornstein-Uhlenbeck process, describing a particle in an external force field $f(x)$. It turns out there are nontrivial symmetries only if $f(x)$ is at most linear. We fully discuss the isotropic case, while for the non-isotropic we only deal with a generic situation (to be defined in detail in the text).
\end{abstract}

\label{firstpage}

%%%% The Article text starts here

%\newpage

%\tableofcontents

%\newpage

\section{Introduction}

Symmetry analysis is since a long time one of the key tools to attack deterministic nonlinear differential equations, both ordinary and partial \gcite{\symmref}; actually, what is nowadays known as Lie theory was created by Sophus Lie precisely to study nonlinear (ordinary) differential equations. The special case escaping an efficient use of Lie theory is that of \emph{Dynamical Systems}, i.e. sets of first order ODEs \gcite{\dsref}.

From the point of view of Physics, there is a class of Dynamical Systems which has a special status, i.e. that corresponding to \emph{Newtonian Mechanics of point particles in a force field}, possibly with dissipation:
\beql{eq:N1} \begin{cases} \dot{x}^i \ = \ v^i & \\ \dot{v}^i \ = \ (1/m) \ F^i (\xb,\vb,t ) & \end{cases} \ . \eeq
In this case one actually sets the equations as a second order system,
\beql{eq:N2} \ddot{x}^i \ = \ (1/m) \ F^i (\xb,\dot{\xb},t) \ , \eeq which also allows for an efficient use of Lie theory.

These equations do generally admit no symmetry, and when $F$ is time autonomous they are generically invariant under time translation only; but for certain forms of $F$, other symmetries can arise. (In the frictionless case and for $F$ arising from a potential $\Phi (\xb)$, these symmetries immediately lead -- through a variational formulation in Lagrangian terms and Noether theorem -- to conserved quantities.)

The description of a physical system as an isolated system only subject to a given external (maybe potential) force and with no dissipation is of course in many -- if not all -- physical situations only an idealization of a more complex situation, in which dissipation is present, and other forces beside those taken into account are also present. In many cases, dissipation can be described in terms of a friction force (not necessarily linear, but however depending on the velocity $\vb$), and the extra forces can be described in terms of a \emph{stochastic perturbation}.

In other words, the ideal system \eqref{eq:N1} should be replaced by the system of \emph{Ito stochastic differential equations} \gcite{\sderef} depending on independent Wiener processes $w^i (t)$,
\beql{eq:NI1} \begin{cases} d x^i \ = \ v^i \, d t \ , &  \\ d v^i \ = \ (1/m) \ \[ F^i  (\xb) \ - \ \la \, v^i \] d t \ + \ \s \, d w^i \ . &  \end{cases} \eeq
This is also known as the \emph{Ornstein-Uhlenbeck} process for a point particle of mass $m$ in the force field $F(\xb)$ \gcite{NelsonBM,Wax}.

It should be noted that from the point of view of Ito equations, this is a \emph{degenerate} system, in that the associated diffusion matrix is in this case a degenerate one,
\beql{eq:Dmat} D \ = \ \s \ \begin{pmatrix} 0 & 0 \\ 0 & I \end{pmatrix} \ . \eeq

Symmetry of Stochastic Differential Equations (SDEs) has been studied only in relatively recent years \gcite{\stochsymmref}. By now, a sound formulation of the theory is available, and we know how to use symmetries to integrate -- or reduce the order of -- SDEs \gcite{Koz1,Koz2,Koz3,GS17,GGPR,GL1,GL2,Koz18a,Koz18b,KozB}. In studying symmetries of SDEs, it is customary to assume that the diffusion matrix has full rank; this already shows that the equations \eqref{eq:NI1} may have special properties from this point of view.

The question we want to tackle in this note, at least for the simplest case of homogeneous and isotropic dissipation and homogeneous and isotropic (but we will also consider the ``generic'' -- in a sense to be specified later on -- anisotropic case) stochastic perturbation, is that of classifying the symmetries of the stochastic system \eqref{eq:NI1} with diffusion matrix \eqref{eq:Dmat}.

We will assume the reader to be familiar with the basics of Lie theory for (deterministic) differential equations \gcite{\symmref}, and recall the relevant concepts and formulas concerning symmetry of stochastic differential equations in Sect.\ref{sec:symmstoch} below, referring to the literature for further detail.

It is appropriate to mention here some general matters concerning the notation we will use in this work.

As customary in discussing symmetries, we will only consider \emph{continuous} ones (i.e. Lie symmetries); Lie symmetries are in fact the only ones which can be used to integrate or reduce the equations. By a standard abuse of notation we will routinely  denote as ``symmetries'' the infinitesimal generators of these.

By ``an equation'', we will always mean possibly -- and actually, generally -- a vector one, i.e. a system of coupled scalar equations. Summation over repeated indices will always be understood. We will also use the shorthand notation
\beq \pa_i \ := \ \frac{\pa}{\pa x^i} \ \ ; \ \ \ \^\pa_k \ := \ \frac{\pa}{\pa w^k} \ \ ; \ \ \ \pa_t \ := \ \frac{\pa}{\pa t} \ . \eeq

A number of Remarks will address side questions, which can be safely skipped on a first reading.

\medskip\noindent
{\bf Remark 1.}
It should be noted that other forms of the noise terms could be considered: e.g. in the context of population dynamics, it would be natural to have noise coefficient proportional to $|x|$ (environmental noise) or to $\sqrt{|x|}$ (demographical noise), as discussed in \gcite{Glogistic}. Our choice \eqref{eq:Dmat} is the natural one in view of the Mechanical origin of the equation \eqref{eq:NI1}. \EOR

\medskip\noindent
{\bf Remark 2.} As well known, to any Ito equation corresponds a Stratonovich equation; the correspondence between the two has actually some subtle point \gcite{Stroock}. There is also, as should be expected, a correspondence between symmetries of an Ito equation and those of the corresponding Stratonovich equation \gcite{Unal,Koz2012,Ikeda,GL1,Koz18b}; see in particular \gcite{Koz2012}, also for correction to previous literature. Here we will only work in the Ito framework, but a completely equivalent (up to the subtle points mentioned above) Stratonovich formulation would also be possible. \EOR

\subsection*{Acknowledgements}

\addcontentsline{toc}{subsection}{Acknowledgements}

I warmly thank an extremely careful and patient Referee. This work was performed at SMRI, providing good working conditions and a relatively relaxed atmosphere despite and through the various degrees of lockdown. My work is also supported by GNFM-INdAM.

\section{Symmetry of stochastic differential equations}
\label{sec:symmstoch}

Let us consider a general Ito equation
\beql{eq:Ito} d x^i \ = \ f^i (x,t) \, d t \ + \ \s^i_{\ k} (x,t) \, d w^k \ , \eeq
where $w^k = w^k (t)$ are standard independent Wiener processes \gcite{\sderef}, and $i,k=1,...,n$.

In the applications of symmetries to integration of the Ito equation, one is mainly interested in \emph{simple symmetries}, i.e. in those not affecting time; the reason for this lies in Kozlov approach \gcite{Koz1,Koz2,Koz3} connecting simple symmetries to integration -- or at least reduction \gcite{GL2} -- of stochastic equations.

In this case the most general generator of (continuous) symmetries reads
\beql{eq:X} X \ = \ \vphi^i (x,t;w) \, \frac{\pa}{\pa x^i} \ + \ h^k (x,t;w) \, \frac{\pa}{\pa w^k} \ = \ \vphi^i \, \pa_i \ + \ h^k \, \^\pa_k \ . \eeq

As discussed in detail in \gcite{GW18} (see in particular Section VI and Lemma 1 in there), the functional form of the $h^i$ is actually strongly constrained: it results that
\beq h^i (x,t;w) \ = \ R^i_{\ j} \, w^j \ , \eeq
with $R$ a matrix in the Lie algebra of the \emph{linear conformal group}, \beql{eq:LCG} O(n) \times ({\bf R}_+)^n \ ; \eeq in practice, $R$ is the sum of a diagonal and a skew-symmetric matrices,
\beql{eq:Rform} R \ = \ D \ + \ S \ , \ \ \ D_{ij} = r_i \de_{ij} \ , \ \ \ S_{ji} = - S _{ij} \ . \eeq
Thus, with the shorthand notation introduced above, we have to consider symmetry generators of the form
\beql{eq:XW} X \ = \ \vphi^i (x,t;w) \, \pa_i \ + \ R^k_{\ m} w^m \^\pa_k \ , \eeq
with $R$ as specified by \eqref{eq:Rform}.

\medskip\noindent
{\bf Remark 3.} Actually the most general acceptable action on the time coordinate is a reparametrization of this, see again \gcite{GW18}, or \gcite{GGPR,GL1,GS17}; this would correspond to a term $\tau (t) \pa_t$ being inserted in $X$. Thus very little is lost in restricting to simple symmetries even from the point of view of classifying general symmetries -- beside the fact that, as already mentioned, non-simple symmetries cannot be used in the Kozlov integration scheme \gcite{Koz1,Koz2,Koz3,GL2}. \EOR

\medskip\noindent
{\bf Remark 4.} Let us recall some convention used in the literature. Symmetries with $R=0$ and $\vphi$ not depending on $w$ are called \emph{deterministic symmetries}; those with $R=0$ but with at least some $\vphi$ depending on some $w$ are called \emph{random symmetries}. Symmetries with $R\not= 0$ are also called \emph{W-symmetries} to emphasize that they affect the Wiener processes\footnote{The condition on $R$ seen above descends from the requirement to map the independent Wiener processes into independent Wiener processes; the effect of the conformal factor ${\bf R}_+$ can be absorbed into the diffusion coefficients $\s$ and this is hence allowed. See \gcite{GS17,GW18} for details.}. It should be stressed that while the previous classes of symmetries are special cases, W-symmetries are actually the most general case, and the name serves to stress that ``full advantage is taken'' of all the possible dependencies -- and hence of all possible transformations. {In the following we will find convenient to have a collective name for all symmetries but W-symmetries; we will refer to deterministic and random symmetries as \emph{regular} symmetries.} \EOR
\bigskip

As shown in Lemma 3 of \gcite{GW18}, the \emph{determining equations} for general (hence W) symmetries of the Ito equation \eqref{eq:Ito} are
\begin{eqnarray}
\pa_t \vphi^i \ + \ \( f^j \, \pa_j \vphi^i \ - \ \vphi^j \, \pa_j f^i \) \ + \ \frac12 \ \Delta \vphi^i &=& 0 \ , \label{eq:deteq1} \\
\^\pa_k \vphi^i \ + \ \( \s^j_{\ k} \, \pa_j \vphi^i \ - \ \vphi^j \, \pa_j \s^i_{\ k} \) \ - \ \s^i_{\ m} \, R^m_{\ k} &=& 0 \ . \label{eq:deteq2} \end{eqnarray}
Here and below, $\Delta$ is the \emph{Ito Laplacian} \gcite{\sderef}
\beql{eq:Lapl} \Delta \ := \ \sum_{j,k} \[ \delta^{jk} \, \frac{\pa^2}{\pa w^j \pa w^k} \ + \ 2 \, \s^{jk} \, \frac{\pa^2}{\pa x^j \pa w^k} \ + \ \( \s \, \s^T \)^{jk} \, \frac{\pa^2}{\pa x^j \pa x^k} \] \ . \eeq
At some point it will be convenient to have a standard notation for referring to these equations; we will then refer to the $i$-th equation in \eqref{eq:deteq1} as $\E^i$, and to the equation with indices $i,k$ in \eqref{eq:deteq2} as $E^i_{\ k}$.

\medskip\noindent
{\bf Remark 5.} The equations \eqref{eq:deteq1} are a ``first block'' of $n$ equations, while \eqref{eq:deteq2} are a ``second block'' of $n^2$ equations. Note that the latter only depend on the diffusion coefficients $\s^i_{\ j}$, but not on the drift coefficients $f^i$; equations in the first block depend on $\s$ through the Ito Laplacian. It will thus be generally convenient to study the equations starting from those in the ``second block''. We will also refer to equations in the two blocks as the ``$f$-determining equations'' and the ``$\s$-determining equations'', respectively, albeit - as mentioned above - this is not completely correct (as the $f$-determining equations also depend on $\s$ through the Ito Laplacian). \EOR

\medskip\noindent
{\bf Remark 6.} Examples of the determination and applications of W-symmetries for Ito equations are provided in \gcite{GW18}; see also \gcite{GS17} for deterministic and random symmetries. \EOR

\medskip\noindent
{\bf Remark 7.} As recalled above, the theory could also be developed in terms of Stratonovich equations; the determining equations for W-symmetries of a Stratonovich equation are also discussed in \gcite{GW18}; see also \gcite{GL1,GL2,GS17} for their deterministic and random symmetries. \EOR

\medskip\noindent
{\bf Remark 8.} W-symmetries have been used to integrated the logistic equation with multiplicative noise (which has a relevant role in Mathematical Biology) \gcite{Glogistic}; the symmetries of scalar equations with multiplicative noise have been completely classified \gcite{GSclass}; this is equivalent to the  classification of equations in such a class which can be integrated by the Kozlov approach \gcite{Koz1,Koz2,Koz3,KozB}. \EOR

\section{Invariants of stochastic differential equations}
\label{sec:invariants}

The work of R. Kozlov \gcite{Koz18a,Koz18b,KozB} stresses the relevance of \emph{invariants} for Ito equations. Given the equation \eqref{eq:Ito}, which will now be denoted as $E$, we say that
$$ \Theta \ = \ \Theta (\xb,t;\wb) $$ is an invariant for the equation if the differential of $\Theta$ computed along the equation vanishes. We have, with standard (Ito) calculus,
\begin{eqnarray*}
d \Theta \vert_E &=& \[ \( \frac{\pa \Theta}{\pa x^i} \) \, d x^i \ + \ \( \frac{\pa \Theta}{\pa w^k} \) \, d w^k \ + \  \[ \( \frac{\pa \Theta}{\pa t} \) \ + \ \frac12 \, \Delta (\Theta) \] \, d t \]_E  \\
&=& \( \frac{\pa \Theta}{\pa x^i} \) \, \( f^i \, dt \ + \ \s^i_{\ k} d w^k \)  \ + \  \( \frac{\pa \Theta}{\pa w^k} \) \, d w^k \ + \  \( \frac{\pa \Theta}{\pa t} \) \, d t \\ & & \  + \ \frac12 \, \[ \Delta (\Theta) \] \, d t \ . \end{eqnarray*}
Thus the requirement $d \Theta \vert_E = 0$ amounts to $n+1$ equations,
\begin{eqnarray}
& & \( \frac{\pa \Theta}{\pa x^i} \) \, \s^i_{\ k} \ + \  \( \frac{\pa \Theta}{\pa w^k} \) \ = \ 0 \ \ \ \ (k=1,...,n) \ ; \\
& & \( \frac{\pa \Theta}{\pa x^i} \) \, f^i \ + \  \( \frac{\pa \Theta}{\pa t} \) \ + \ \frac12 \, \[ \Delta (\Theta) \] \ = \  0 \ . \end{eqnarray}

If the equation \eqref{eq:Ito} admits an invariant $\Theta$, then the invariant  will not change along the evolution described by the equation; it is thus obviously convenient to change variables $(\xb,\wb)$, taking $\Theta$ to be one of the new variables.

It should be noted that such a change of variables will in general be possible only locally (due to singularities of the Jacobian); a relevant exception is obtained when $\Theta$ is a linear function of its $(x^i,w^k)$ -- or at least of the $x^i$ --  arguments.

\medskip\noindent
{\bf Example \ref{sec:invariants}.1.} \gcite{Koz18a} Consider the geometric Brownian motion equation (here $\a,\b$ are real constants)
$$ d x \ = \ \a \, x \, d t \ + \ \b \, x \, d w \ ; $$
for this the function
$$ \Theta \ = \ x \ \exp \[ - \, \( \a - \frac12 \b^2 \) \, t \ - \ \b \, w (t) \]  $$
is an invariant. \EOR

\section{The algebraic structure of symmetries of an Ito equation}
\label{sec:structure}

It was shown in \gcite{Koz18a} that:

\medskip\noindent
{\bf Proposition \ref{sec:structure}.1.} {\it The commutator of two symmetry generators $X_1,X_2$ for a given Ito equation is still a symmetry generator.}

\medskip\noindent
{\bf Proof.} This is shown by direct computation in \gcite{Koz18a}. The same conclusion can be reached by recalling the correspondence between symmetries of an Ito equation and of the corresponding Stratonovich equation. The structure of the determining equations for symmetries of a Stratonovich equation shows immediately that the commutator of (vector fields having as coefficients) two solutions is still a (vector field having as coefficients) a solution. \EOR

\medskip\noindent
{\bf Corollary \ref{sec:structure}.1.} {\it The symmetry generators of an Ito equation form a Lie algebra.}

\medskip\noindent
{\bf Proposition \ref{sec:structure}.2.} {\it If $X$ is a symmetry generator for the Ito equation, then $Y = \a X$ is a symmetry generator if and only if $\a$ is an invariant for the same Ito equation.}

\medskip\noindent
{\bf Proof.} This is shown also in \gcite{Koz18a}, but we give here a simple direct proof.

In fact, the determining equations for $Y = \psi^i \pa_i$ with $\psi^i = \a \vphi^i$ read
{\small
\begin{eqnarray}
\a \[ \pa_t \vphi^i  +  f^j  \pa_j \vphi^i  - \vphi^j  \pa_j f^i + \frac12  \Delta \vphi^i \]  + \( \pa_t \a + f^j  \pa_j \a  +  \frac12  \Delta \a \)  \vphi^i
{\ + \ \frac12 \ \mathcal{Q} [\a , \vphi^i ] } \ = \ 0 & , & \label{eq:LM1} \\
\a \[ \^\pa_k \vphi^i + \s^j_{\ k} \pa_j \vphi^i - \vphi^j \pa_j \s^i_{\ k} \]   + \( \^\pa_k \a + \s^j_{\ k}  \pa_j \a \)  \vphi^i \ = \ \a \, \s^i_{\ j} R^j_{\ k} & . & \label{eq:LM2} \end{eqnarray} }
{Here we have denoted shortly by $\mathcal{Q}$ the term, arising from $\Delta (\a \vphi^i )$, containing first order derivatives; this is given explicitly by
\begin{eqnarray} \mathcal{Q} [\a , \vphi^i ] & := &  (\^\pa_k \a ) (\^\pa_k \vphi^i ) \ + \ \s^{jk} \[ (\^\pa_k \a) (\pa_j \vphi^i ) \, + \, (\pa_j \a) (\^\pa_k \vphi^i) \] \nonumber \\
 & & \ + \ \s^{jk} \s^{\ell k} (\pa_j \a) (\pa_\ell \vphi^i) \ . \label{eq:mathcalQ}\end{eqnarray}}
For deterministic or random symmetries (i.e. regular ones, see Remark 4), we have $R=0$, i.e. the r.h.s. of the second set of equations is identically zero.

If $X = \vphi^i \pa_i$ is a symmetry generator, the term in square bracket in \eqref{eq:LM1} is zero, and the term in square bracket in \eqref{eq:LM2} is just $\s R$. On the other hand, if $\a$ is an invariant, the terms in round brackets are zero.

{Thus we only have to show that $\mathcal{Q} [\a , \vphi^i] = 0$. This follows at once rewriting \eqref{eq:mathcalQ} as
\begin{eqnarray*} \mathcal{Q} &=&  \[ (\^\pa_k \a) \, \( \^\pa_k \vphi^i \ + \ \s^{jk} \, \pa_j \vphi^i  \) \ + \ \s^{jk} \, (\pa_j \a) \, \( \^\pa_k \vphi^i \, + \, \s^{\ell k} \, \pa_\ell \vphi^i  \) \] \\
&=& \( \^\pa_k \a \ + \ \s^{jk} \, \pa_j \a \) \ \( \^\pa_k \vphi^i \ + \ \s^{\ell k} \, \pa_\ell \vphi^i \) \ ; \end{eqnarray*}
again, the term in the first bracket vanishes since $\a$ is assumed to be an invariant.} \EOR

{
\medskip\noindent
{\bf Remark 9.} There may seem to be a contradiction between assuming the entries of $R$ to be constant and considering the vector field $Y = \a X$ with $\a$ a function. But this function is by definition a constant on the dynamic defined by the Ito equation, so the entries of $\a R$ are also constant on this dynamics. \EOR}

\medskip\noindent
{\bf Corollary \ref{sec:structure}.2.} {\it The symmetry generators of an Ito equation have, beside the structure of Lie algebra, also that of a \emph{Lie module}.}

\medskip\noindent
{\bf Remark 10.} These properties correspond to those for symmetries of a deterministic dynamical system \gcite{CGbook}. \EOR

\medskip\noindent
{\bf Remark 11.} The discussion in Sect.IX of  \gcite{GS17} seems to be in contradiction to this. The reason is that in there the restriction on the functional form of $h(\xb,t;\wb)$ introduced in Sect.\ref{sec:symmstoch}, i.e. $h (\xb,t;\wb) = R \wb$, was not implemented (as the need for this restriction was only discussed in the later paper \gcite{GW18}). Also, when one implements this restriction in Example 11 of \gcite{GS17}, it turns out no simple $t$-independent symmetries exist. \EOR

\section{Integration of stochastic equations via symmetry and/or invariants}
\label{sec:integr_reduct}

The relevance of symmetries in the analysis of SDEs lies in that once (and if) symmetries are determined, they can be used \emph{constructively} to integrate the equation, or at least to reduce it to a lower dimensional one.

\subsection{Integration or reduction via invariants}

Invariants can be readily used to express (at least locally, as already remarked) the solution $x(t)$ of a SDE in terms of the invariants themselves -- and of course of the realization of the involved Wiener processes.

Note that if a higher dimensional equation has a number of invariants smaller than its dimension, the reduction by invariants will in general only allow to express some of the components $x^i (t)$ of the solution in terms of the invariants and of the other components.

\medskip\noindent
{\bf Example \ref{sec:integr_reduct}.1.} \gcite{Koz18a} For the geometric Brownian motion considered in the Example \ref{sec:invariants}.1 above, knowledge of the invariant $\Theta$ immediately allows to write $x(t)$ in the form $$ x (t) \ = \ C \ \exp \[ \( \a - \frac12 \b^2 \) \, t \ + \ \b \, w (t) \] \ , $$ with $C = \Theta (0)$ an arbitrary constant (in fact, given by the value of $\Theta$ at the initial time $t=0$; for $w(0)=0$ this is just $x(0)$). This provides a full solution to our stochastic equation, giving an explicit expression for $x(t)$ for each realization of the driving Wiener process $w(t)$. \EOR

\subsection{Integration or reduction via symmetries}

We can now consider the use of symmetries for solving, or at least reducing, stochastic equations.
Let us first consider the scalar case, in which determination of a single \emph{simple} symmetries allows to integrate the equation.

The key observation is that (denoting for a moment the spatial variable as $y$ and the Wiener process as $z(t)$) if we have a symmetry of the form
\beql{eq:X0K} X \ = \ \pa_y \ , \eeq
then necessarily the Ito equation is of the special form
\beql{eq:Itospecial1d} d y \ = \ f(t) \, d t \ + \ \s (t) \, d z \ . \eeq

In fact, consider the determining equations \eqref{eq:deteq1}, \eqref{eq:deteq2}, seeing these now as equations for the $f$ and $\s$ coefficients with $\phi = 1$ and $R=0$  given (vector indices are absent as we deal with the scalar case). They are just
$$ \pa_y f \ = \  0 \ , \ \ \ \pa_y \s \ = \ 0 \ . $$

Now, the point is that eq.\eqref{eq:Itospecial1d} is promptly integrated, and we get
\beq y(t) \ = \ y(t_0) \ + \ \int_{t_0}^t f(t) \, d t \ + \ \int_{t_0}^t \s(t) \, d z(t) \ . \eeq

Thus, if we determine a symmetry of the general form \eqref{eq:X} for the Ito equation \eqref{eq:Ito} {(which, we recall, is in this subsection specialized to the scalar case)}, we can seek a change of coordinates $(x,t;w) \to (y,t;z)$ mapping this {symmetry} into \eqref{eq:X0K}; as symmetries are preserved under diffeomorphisms \gcite{GL1}, in the new coordinates the equation will be in the form \eqref{eq:Itospecial1d}, and thus promptly integrated. It should be stressed that symmetry is not only a sufficient condition for integrability \gcite{Koz1,Koz2,Koz3}, but a necessary one as well \gcite{GL2}.

The situation is only slighter more complicated, but not conceptually different, if we deal with higher dimensions and with higher dimensional symmetry algebras \gcite{GL2}.
Essentially, reduction will be performed by stages, through multiple changes of coordinates; at each stage one of the scalar equations can be integrated in terms of the solutions to the remaining system. Note that, as always in symmetry analysis, multiple reduction should be performed following the algebraic structure \gcite{\symmref}, or one could be unable to take full advantage of the symmetries (these could be ``lost in reduction'' if the proper order is not followed).

We give here a very simple example in dimension one. Examples with higher dimensional equations or with random or W-symmetries would be more involved; we refer the reader to the literature, see in particular \gcite{Koz2,GL2,GGPR,GS17,GW18}, for these.

\medskip\noindent
{\bf Example \ref{sec:integr_reduct}.2.} \gcite{GL1,GL2} The Ito equation
\beql{eq:example1} d y \ = \ \[ e^{- y} \ - \ (1/2) \, e^{-2 y} \]
\, d t \ + \ e^{- y} \, d w \eeq admits the vector field $ X  =
e^{- y} \pa_y $ as a symmetry generator. By the change
of variables $x = \exp[y]$ the vector field reads $X = \pa_x$, and
the initial equation \eqref{eq:example1} reads
$$ d x \ = \ d t \ + \ d w \ . $$
Thus we have $x(t)  =  x(t_0)  +  (t - t_0 )  +  [w(t) - w(t_0)]$ (note this is positive for $x(t_0)>1$).
Inverting the change of coordinates, $y(t) = \log [x(t)]$, we get a solution to the original equation. \EOR

\medskip\noindent
{\bf Remark 12.} It should be stressed that the application to the case of deterministic or random symmetries is rather straightforward (see e.g. Sect.7 in \gcite{GL2}): the required change of variables does not affect neither time nor the Wiener processes and in the scalar case it is just $x = \Phi (y,t;w)$ with \beq \Phi (y,t;w) \ = \ \int \frac{1}{\phi (x,t;w)} dy \ . \eeq
On the other hand, in the case of W-symmetries we are not guaranteed that operating with these changes of symmetry we remain within the framework of Ito equations: thus, albeit the procedure is exactly the same, we will discover only \emph{a posteriori} if it provides an integration of the equation under study.\footnote{This question could also be studied \emph{a priori}, i.e. without actually implementing the change of coordinates; but in practice this is harder than just changing coordinates and check if we are in the favorable case. See \gcite{GW18} for details.} \EOR

\section{Ito equations for an isotropic Ornstein-Uhlenbeck process}
\label{sec:Ito_OU}

We should now consider the dynamical system describing the motion of a particle (or system of particles) in a force field which is in part due to a potential and in part due to a stochastic perturbation -- as in Brownian Motion.

It is well known that it does not suffice to add a fluctuating force term: in the case of Brownian motion this follows from the fact that a particle with a net drift in some direction collides with more background particles on the front side than on the back side; but it is also needed in general to keep the average (in statistical sense) Energy constant.

Thus a classical particle (of unit mass, for ease of notation) moving in a force field ${\bf F} (\xb , t)$ (possibly, but not necessarily, originating from a potential and subject moreover to a (constant intensity) stochastic force obeys the Ito equation
\begin{eqnarray}
d x^i &=& v^i \, dt \nonumber \\
d v^i &=& \[ F^i (\xb,t) \ - \ \b \, v^i \] \, dt \ + \ \s \, d w^i \label{eq:BM}  \end{eqnarray}
with $\b$ and $\s$ positive real constants.

\medskip\noindent
{\bf Remark 13.} For the physical Brownian Motion these constants are classically related to each other -- and to the temperature $T$ -- by
\beql{eq:sigmabeta} \s^2 \ = \  2 \ \frac{\b \, \kappa \, T}{m} \ , \eeq
where $\kappa$ is the Boltzmann constant, and $T$ the absolute temperature. See e.g. \gcite{NelsonBM} for a careful discussion of the derivation of the equations \eqref{eq:BM}. \EOR
\bigskip

The equations \eqref{eq:BM} can be generalized by introducing anisotropic and possibly inhomogeneous friction and underlying stochastic perturbation terms. Considering these would be conceptually non different from the isotropic case, but would introduce some (nontrivial) algebraic complications. We will thus stick to the isotropic case, albeit we will start by discussing the general -- hence possibly anisotropic -- case in Sect.\ref{sec:general}.

As mentioned above, we can consider not only the case where the deterministic forces arise from a potential, but also the more general case of a force field, and the case where even in the absence of an underlying Brownian field the particle is experiencing a friction drag. In this sense the coefficient $\b$ should not necessarily be of the form required by \eqref{eq:sigmabeta}.

\medskip\noindent
{\bf Remark 14.} We stress again that when referring to the \emph{isotropic case}, we always mean that the friction and diffusion coefficients $\b$ and $\s$ are the same for the different degrees of freedom, and correspondingly in the \emph{anisotropic case} we drop this assumption. No reference is made to the external force field ${\bf F} (\xb)$, which even in the isotropic case can very well be anisotropic. \EOR

%\newpage

\section{Symmetries and invariants for the Ornstein-Uhlenbeck process. General features}
\label{sec:general}

Our discussion will show that results depend on the form of the external force field, in particular on the degree of the function $\fb (\xb)$, and on the system being isotropic or anisotropic in the sense discussed at the end of the previous Section.

In this Section we will discuss, as far as possible, the situation for invariants and symmetries in general terms, i.e. for a system with $n$ degrees of freedom -- which gives a system of Ito equations of dimension $2 n$. We will arrive at a set of reduced determining equations and a reduced general form for the symmetry generator; further discussion will depend on the aforementioned details, and is postponed to ensuing Sections.

Our general formalism, see Sect.\ref{sec:symmstoch}, would require to consider $2 n$ independent Wiener processes; but it is clear from \eqref{eq:BM} that only $n$ of these appear in the equations. We will thus denote these $2 n$ Wiener processes as $\{ z^1,w^1,z^2,w^2,...,z^n,w^n\}$; the $w^i$ do play a role in the equations and will play a real role in our discussion, while the $z^i$ are ``ghost'' Wiener processes.

Correspondingly, the matrix $\s$ will be written as
\beq \s \ = \ \mathrm{diag} \( 0 , \mu_{(1)} , ... , 0 , \mu_{(n)} \) \ , \eeq which of course yields
\beq \s \ \s^T \ = \ \mathrm{diag} \( 0 , \mu_{(1)}^2 , ... , 0 , \mu_{(n)}^2 \) \ . \eeq

\subsection{Invariants}
\label{sec:gen_inv}

As discussed in Sect.\ref{sec:symmstoch}, knowledge of invariants can be helpful in analyzing symmetries, and sometimes even in determining solutions to a stochastic equation. We will thus begin with studying invariants for the equation \eqref{eq:BM}.

In general terms, we will consider a scalar function
\beq h \ = \ h (\xb , \vb , t ; \zb , \wb ) \ . \eeq
We will then consider its differential
$$ d h \ = \ \frac{\pa h}{\pa x^i} \, d x^i \ + \ \frac{\pa h}{\pa v^i} \, d v^i \ + \ \frac{\pa h}{\pa z^i} \, d z^i \ + \ \frac{\pa h }{\pa w^i} \, d w^i \ + \ \[ \frac{\pa h}{\pa t} \ + \ \frac12 \, \Delta (h) \] \, dt $$
and evaluate it on solutions to \eqref{eq:BM}, i.e. substitute for $dx^i$ and $d v^i$ according to this; in this way we will obtain an expression in $d t , dz^i , dw^i$. For the latter to vanish, all coefficients of the different differentials must vanish separately.

Proceeding in this way, and considering first the coefficient of $d z^i$, we immediately obtain that
\beq \frac{\pa h}{\pa z^i} \ = \ 0 \ . \eeq
Moreover, considering also the coefficient of $d w^i$, we obtain that
\beq h \ = \ P (\xb , t ; \zetab ) \eeq
having defined the characteristic variables
\beq \zeta^i \ := \ w^i \ - \ \frac{v^i}{\mu_{(i)}} \ . \eeq
Note that we are performing a change of variables $(\xb,\vb,t;\zb,\wb) \to (\xb,\vb,t;\zb,\zetab)$.

Now $dh = dP$ contains only the $dt$ differential, i.e. we are reduced to a single equation. In this, however, we have terms which are linear in the $v^i$. As the unknown function $P$ does not depend on these variables, their coefficient on the equation must vanish separately. Such coefficients are of the form
\beq \frac{1}{\mu_{(1)} ... \mu_{(n)}} \ \[ \frac{\b_{(i)}}{\mu_{(i)}} \, \frac{\pa P}{\pa \zeta^i} \ + \ \frac{\pa P}{\pa x^i} \] \ , \eeq hence their vanishing imply that
\beq P (\xb , t ,\zetab ) \ = \ Q (\ub , t ) \ , \eeq
having defined the new characteristic variables
\beq u^i \ := \ \zeta^i \ - \ \frac{ \b_{(i)} }{ \mu_{(i)} } \ x^i \ . \eeq
Introducing these, we are considering a second change of variables
\beql{eq:inv2cv} (\xb,\vb,t;\zb,\zetab) \ \to \ (\xb,\vb,t;\zb,\ub) \ . \eeq
With this, we reach the expression
\beq d h = d Q \ = \ \frac{\pa Q}{\pa t} \ - \ \[ \sum_{i=1}^n \frac{F^i (\xb)}{\mu_{(i)}} \ \frac{\pa Q}{\pa u^i} \] \ . \eeq
That is, invariants are identified by solutions to
\beql{eq:Qinv} \frac{\pa Q}{\pa t} \ = \ \[ \sum_{i=1}^n \frac{F^i (\xb)}{\mu_{(i)}} \ \frac{\pa Q}{\pa u^i} \] \ , \ \ \ \ Q = Q (\ub,t) \ . \eeq

Obviously, solutions to this equation will depend on the assigned functions $F^i (\xb)$ and on the assigned nonzero real constants $\mu_{(i)}$. We cannot go any further in full generality. We note however two special cases.

\subsubsection{Constant force}

First of all, suppose that $\fb (\xb) = {\bf c}$. Then, writing $\rho^i := c^i/\mu_{(i)}$, equation \eqref{eq:Qinv} reads
\beql{eq:QinvC} \frac{\pa Q}{\pa t} \ = \ \[ \sum_{i=1}^n \rho^i \ \frac{\pa Q}{\pa u^i} \] \eeq and is solved by an arbitrary function
\beq Q \ = \ Q (\chi^1 , ... , \chi^n ) \eeq
of the characteristic variables
\beql{eq:chirho} \chi^i \ := \ u^i \ + \ \rho^i \, t \ . \eeq
We have thus shown that:

\medskip\noindent
{\bf Lemma \ref{sec:general}.1.} {\it The the $\chi^1 , ... \chi^n$ are the basic invariants for the equation \eqref{eq:BM} with constant (possibly zero) external forces $F^i (\xb) = c^i$.}

\subsubsection{Linear force}

In the case where $F^i (x)$ is a linear function,
\beq F^i (\xb) \ = \ L^i_{\ j} \, x^j \ + \ K^i \ , \eeq
the equation \eqref{eq:Qinv} decouples into $n+1$ equations.

The terms independent of the $x$ yield again \eqref{eq:QinvC} but now with
\beq \rho^i \ := \ K^i / \mu_{(i)} \ . \eeq
(Note the equation reduces to $\pa Q / \pa t = 0$ for $K^i = 0$.)

The terms in which $x^i$ appears yield the equation
\beql{eq:FQG} \sum_{j=1}^n \frac{1}{\mu_{(j)}} \ \frac{\pa Q}{\pa u^j} \ L^j_{\ i} \ = \ 0 \ . \eeq

This set of $n+1$ PDEs does not admit a nontrivial solution when $L$ is non-degenerate. In fact, define the matrix $\wt{L}$ with entries
\beq \wt{L}^i_{\ j} \ = \ \mu_{(i)}^{-1} \ L^i_{\ j} \ ; \eeq we have
$$ \det (\wt{L} ) \ = \ \frac{1}{\mu_{(1)} ... \mu_{(n)} } \ \det (L) $$ so $L$ regular implies that $\wt{L}$ is also regular, and acting on \eqref{eq:FQG} by $\wt{L}^{-1}$ we obtain that necessarily $(\pa Q / \pa u^i) = 0$. In view of \eqref{eq:Qinv} this also implies that $(\pa Q / \pa t) = 0$, so we are left with trivial invariants only:

\medskip\noindent
{\bf Lemma \ref{sec:general}.2} {\it The equation \eqref{eq:BM} with linear external forces $F^i (\xb) = L^i_{\ j} x^j$ and $L$ non-degenerate, admits no non-trivial invariant.}

\subsection{Symmetries}
\label{sec:gen_sym}

We come now to considering symmetries of the general Ornstein-Uhlenbeck process \eqref{eq:BM}. We will write a general vector field (not acting on $t$, as implied by our choice of considering only simple symmetries) in the form
\beq X \ = \ \sum_{i=1}^n \xi^i (\xb,\vb,t;\zb,\wb) \, \frac{\pa}{\pa x^i} \ + \ \eta^i (\xb,\vb,t;\zb,\wb) \, \frac{\pa}{\pa v^i} \ + \ X_R \ , \eeq where
\beql{eq:XRrev} X_R \ = \ \(R^{2i-1}_{\ 2j-1} z^j \ + \ R^{2i - 1}_{\ 2 j} w^j \) \, \frac{\pa}{\pa z^i} \ + \ \(R^{2i}_{\ 2j-1} z^j \ + \ R^{2i }_{\ 2 j} w^j \) \, \frac{\pa}{\pa w^i} \eeq is the part related to the $R$ matrix and present only in the proper W-symmetries.

We proceed then to writing the determining equations \eqref{eq:deteq1}, \eqref{eq:deteq2}. As already remarked, it is convenient to start analyzing the ``second block'', i.e. the $\s$-determining equations \eqref{eq:deteq2}.

\subsubsection{The $\s$-determining equations}

The equations $E^{2i-1}_{\ 2j -1}$ read
\beq \frac{\pa \xi^i}{\pa z^j} \ = \ 0 \ , \eeq
which of course tells that the $\xi^i$ do not depend on the ``ghost'' variables $z^j$.

The equations $E^{2i-1}_{\ 2 j}$ read instead
\beq \frac{\pa \xi^i}{\pa w^j} \ + \ \mu_{(j)} \ \frac{\pa \xi^i}{\pa v^j} \ = \ 0 \ . \eeq
Taking account of both these groups of equations (that is, of all equations $E^{2i-1}_{\ k}$ with odd upper index) we have
\beql{eq:xihatN} \xi^i \ = \ \^\xi^i (\xb,t,\zetab) \ , \eeq
having defined the characteristic variables\footnote{{Note that we are (so far implicitly) considering a change of variables; the new variables $\zeta^i$ will replace, in our list of variables, either the $v^i$ or the $w^i$. We will make our choice later on.}}
\beql{eq:zetacv} \zeta^i \ := \ w^i \ - \ \frac{v^i}{\mu_{(i)}} \ . \eeq

{Having solved the equations $E^{2i-1}_k$ with odd upper index, let us come to the equations $E^{2i}_k$ with even upper index. These do not involve the functions $\xi^i$ and we will thus start operating on them with the original variables $(\xb,\vb,t,\zb,\wb)$.}

The equations $E^{2i}_{\ 2j-1}$ read
\beq \frac{\pa \eta^i}{\pa z^j} \ = \ \mu_{(i)} \ R^{2i}_{\ 2 j - 1} \ ; \eeq
these do of course imply
\beq \eta^i \ = \ \wt{\eta}^i (\xb,\vb,t;\wb) \ + \ \mu_{(i)} \ R^{2i}_{\ 2j-1} \ z^j \ . \eeq
With this, the equations $E^{2i}_{\ 2j}$ read
\beq \frac{\pa \wt{\eta}^i}{\pa w^j} {\ + \ \mu_j \ \frac{\pa \wt{\eta}^i}{\pa v^j} } \ = \ \mu_{(i)} \ R^{2i}_{\ 2j} \ . \eeq
{Solving these leads again to the appearance of the characteristic variables $\zeta^i$ defined above in \eqref{eq:zetacv}. More precisely,} we get
\beq \wt{\eta}^i (\xb,{\vb},t;\wb) \ = \  \^\eta (\xb,t,{\zetab}) \ + \ \mu_{(i)} \ R^{2i}_{\ 2j} \, w^j \ , \eeq which in view of the above means\footnote{{We could of course write the $\eta^i$ in terms of $(\xb,\zetab,t)$ and $\vb$ (instead of $\wb$), but this would give slightly more involved formulas.}}
\beql{eq:etahatN} \eta^i \ = \ \^\eta^i (\xb,t,{\zetab}) \ + \ \mu_{(i)} \ \[ R^{2 i}_{2 j - 1} \, z^j \ + \ R^{2 i}_{2 j} \, w^j \] \ . \eeq

We have thus solved all the $E^i_{\ k}$ equations; the form of $\xi$ and $\eta$ has been restricted to the forms \eqref{eq:xihatN}, \eqref{eq:etahatN}, while $R$ is not restricted yet. Note that it is convenient to use the $\zeta$ variables as new variables; in order to do this we change variables by
$$ (\xb,\vb,t;\zb,\wb) \ \to \ (\xb,t,\zetab;\zb,\wb ) \ . $$

\subsubsection{The $f$-determining equations}

We should now tackle the $f$-determining equations, i.e. the $n$ equations $\E_k$, see \eqref{eq:deteq1}.

These have a rather involved expression which we do not report here. It should be noted that the unknown functions $\^\xi$, $\^\psi$ do \emph{not} depend on the $z$ and $w$ variables; thus all dependencies of these are fully explicit, and the coefficients of different monomials in these should vanish separately.

Actually, the $\E_k$ only contain terms linear in the $z^i$ and $w^i$, which simplifies our task. The vanishing of the terms in $z^m$ in $\E_k$ (or, equivalently, the differential consequence $\pa \E_k / \pa z^m$) yields
$$  \mu_{(k)} \ R^{2 k}_{\ 2m-1} \ = \ 0  \ ; $$ as by assumption the $\mu_{(k)}$ are nonzero, this means that
\beq R^{2k}_{\ 2m-1} \ = \ 0 \ . \eeq Moreover, by the general form of $R$, we should also have (this and the previous equation hold for all $k,m=1,...,n$)
\beq R^{2 m - 1}_{\ 2k} \ = \ 0 \ . \eeq
Thus, the $R$ matrix has nonzero entries $R^i_{\ j}$ only if both indices have the same parity. It is then convenient to consider the reduced $n$-dimensional $R$ matrices $\^R$ and $\wt{R}$, with entries
\beql{eq:RR} \^R^i_{\ j} \ := \ R^{2i}_{\ 2j} \ , \ \ \ \wt{R}^i_{\ j} \ := \ R^{2i-1}_{\ 2j-1} \ . \eeq

Now the equations $\E_k$ do not contain any $z^i$ term, and the $X_R$ vector field defined in \eqref{eq:XRrev} reads
\beq X_R  \ = \ \( \wt{R}^{i}_{\ j} \, z^j \) \ \frac{\pa }{\pa z^i} \ + \ \( \^R^{i}_{\ j} \, w^j \) \ \frac{\pa }{\pa w^i} \ . \eeq

The same argument used for $z^i$ can be used for $w^i$. The vanishing of the term in $w^m$ within $\E_{2k-1}$ (or equivalently the differential consequence $\pa \E_{2k-1} / \pa w^m$) reads
\beq \b_{(m)} \ \frac{\pa \^\xi^k}{\pa \zeta^m} \ + \ \mu_{(m)} \ \frac{\pa \^\xi^k}{\pa x^m} \ = \ \mu_{(k)} \ R^{2 k}_{\ 2 m} \ ; \eeq
these are solved by
\beq \^\xi^i (\xb,t,\zetab ) \ = \ \^\phi (\ub , t) \ + \ \sum_{j=1}^n \frac{\mu_{(i)}}{\mu_{(j)}} \ \^R^{i}_{\ j} \ x^j \ , \eeq
having defined the new characteristic variables
\beql{eq:uN} u^i \ := \ \zeta^i \ - \ \frac{\b_{(i)}}{\mu_{(i)}} \ x^i \ = \ w^i \ - \ \frac{1}{\mu_{(i)}} \, v^i \ - \ \frac{\b_{(i)}}{\mu_{(i)}} \ x^i \ . \eeq

We still have to impose the vanishing of the term in $w^m$ within $\E_{2k}$ (or equivalently the differential consequence $\pa \E_{2k}/\pa w^m$); this reads
\beq \mu_{(m)} \ \frac{\pa \^\eta^k}{\pa x^m} {\ + \ \b_{(m)} \ \frac{\pa \^\eta^k}{\pa \zeta^m} } \ = \ \ - \b_{(k)} \ \mu_{(k)} \ \^R^{k}_{\ m} , \eeq
and enforces
\beq \^\eta^i \ = \ \^\psi^i ({\ub},t) \ - \ \b_{(i)} \ \sum_{j=1}^n \frac{\mu_{(i)}}{\mu_{(j)}} \ \^R^{i}_{\ j} \ x^j \ . \eeq

\medskip\noindent
{\bf Remark 15.}
It may be worth pausing to make the point of the situation: we have reduced the equations $\E_k$ by eliminating all terms linear in the $z^i$ and $w^i$ variables; in doing so we have reached the functional forms 
\begin{eqnarray}
\xi^i &=& \^\phi^i (\ub,t) \ + \ \sum_{j=1}^n \frac{\mu_{(i)}}{\mu_{(j)}} \ \^R^{i}_{\ j} \ x^j \ , \label{eq:redxifin} \\
\eta^i &=& {\^\psi^i (\ub,t)} \ + \ \[ \mu_{(i)} \ \sum_{j=1}^n \^R^{i}_{\ j} \ w^j \ - \
\sum_{j=1}^n \b_{(i)} \ \frac{\mu_{(i)}}{\mu_{(j)}} \ \^R^{i}_{\ j} \ x^j \] \ ; \label{eq:redetafin} \end{eqnarray}
{the characteristic variables $u^i$ are defined in \eqref{eq:uN}.}
As for the matrix $R$, it contains only terms with both indices odd or both indices even; see \eqref{eq:RR} for the relation between $\^R$ and $\wt{R}$ and $R$.

We still have the equations $\E_k$, which albeit reduced are rather involved. In these, the force functions $F^i (\xb)$ appear explicitly. The reduced equations are 
\begin{eqnarray} \frac{\pa \^\phi^i}{\pa t} & - & \sum_{j=1}^n \( \frac{\pa \^\phi^i}{\pa u^j} \) \ \frac{F^j}{\mu_{(j)}} \ = \ \^\psi^i (t) \ + \ \mu_{(i)} \ \( \sum_{j=1}^n \, \^R^{i}_{\ j} \, u^j \) \nonumber \\
& & \ - \ \( \sum_{j=1}^n \frac{\mu_{(i)}}{\mu_{(j)}} \, \( \b_{(i)} - \b_{(j)} \) \ \^R^{i}_{\ j} \, x^j \) \ , \label{eq:redxigen} \\
\frac{\pa \^\psi^i}{\pa t} & - & {\sum_{j=1}^n \( \frac{\pa \^\psi^i}{\pa u^j} \) \ \frac{F^j}{\mu_{(j)}} } \ + \ \b_{(i)} \, \^\psi^i \ = \ \sum_{j=1}^n \( \frac{\pa F^i}{\pa x^j} \) \ \[ \^\phi^j \ + \ \frac{\mu_{(i)}}{\mu_{(j)}} \, \^R^{i}_{\ j} \, x^j \]  \nonumber \\ & & \ - \ \b_{(i)} \, \mu_{(i)} \sum_{j=1}^n \( \^R^{i}_{\ j} \, u^j \) \ + \ \b_{(i)} \sum_{j=1}^n \( \frac{\mu_{(i)}}{\mu_{(j)}} \, \( \b_{(i)} - \b_{(j)} \) \, \^R^{i}_{\ j} \, x^j  \) \ . \label{eq:redetagen} \end{eqnarray}
It is clear that these will simplify in the isotropic case, to be dealt with in the later Section \ref{sec:n_dim}, devoted to it. 
\EOR
\bigskip

As in the discussion about invariants, we note that obviously, solutions to this equation will depend on the assigned functions $F^i(x)$ and on the assigned nonzero real constants $\b_{(i)}$, $\mu_{(i)}$. Thus, here too we cannot go any further in full generality.

\section{The one-dimensional Ornstein-Uhlenbeck process}
\label{sec:1Dsymm}

We will now start considering some special cases, in which the general analysis conducted in Sect.\ref{sec:general} above can be furthered.

The simplest case is of course the one-dimensional one, $n=1$. In this case we just write $\b = \b_{(1)}$, $\mu = \mu_{(1)}$, $x = x^1$, and so on.  In other words, we deal with the equation
\beql{eq:1Deq} \begin{cases} d x \ = \ v \ dt \ , &  \\ d v \ = \ [ F(x) \ - \ \b \ v ] \ dt \ + \ \mu \ d w \ . &  \end{cases} \eeq
We also write, for ease of notation, $R_{11} = \rho$, $R_{22} = r$.

\subsection{Invariants}
\label{sec:lin_invar}

Before analyzing symmetries of \eqref{eq:1Deq}, we will study its invariants. In our simple setting, the equation \eqref{eq:Qinv} reads simply
\beql{eq:Qinv1D} \frac{\pa Q}{\pa t} \ = \ \frac{F(x)}{\mu} \ \frac{\pa Q}{\pa u} \ , \ \ \ \ Q = Q (u,t) \ . \eeq
Differentiating this w.r.t. $x$, we get immediately
$$ F' (x) \ (\pa Q / \pa u) \ = \ 0 \ . $$

Now two possibilities should be considered. If $F' (x) \not= 0$, then we must have $(\pa Q / \pa u) = 0$; this in turn implies, by \eqref{eq:Qinv1D}, $(\pa Q / \pa t) = 0$ as well. That is, in this case no nontrivial invariants exist.

On the other hand, if $F(x) = c$ (with $c$ a constant), then \eqref{eq:Qinv1D} is solved by
\beq Q (u,t) \ = \ \ga [ u + (c/\mu) t ]  \eeq
with $\ga$ an arbitrary function.

In other words, we have the basic invariant
\beql{eq:Jc} J \ = \ u \ + \ (c/\mu ) \ t \ = \ w \ - \ \frac{1}{\mu} \, v \ - \ \frac{\b}{\mu} \, x \ + \ \frac{c}{\mu} \, t \ . \eeq
We summarize the result of our discussion as follows:

\medskip\noindent
{\bf Lemma \ref{sec:1Dsymm}.1.} {\it The equation \eqref{eq:1Deq} admits no nontrivial invariant unless $F(x)=c$; in this case, the ring of invariants is generated by $J$ given in eq.\eqref{eq:Jc}.}

\subsection{Symmetries}

Let us now consider symmetries of \eqref{eq:1Deq}. In this setting, our final results from the discussion in Sect.\ref{sec:general} read 
\begin{eqnarray*}
\xi &=& \^\phi (u , t) \ + \ r \ x \ , \\
\eta &=& \^\psi (u , t) \ + \ \[ \mu \ r \ w \ - \
\b \ r \ x \] \ ;  \\
R &=& \begin{pmatrix} \rho & 0 \\ 0 & r \end{pmatrix}  \end{eqnarray*}
(with $\rho$ an arbitrary constant) for what concerns the symmetry vector components, while the reduced determining equations are \begin{eqnarray} \frac{\pa \^\phi}{\pa t} \ - \ \( \frac{\pa \^\phi}{\pa u} \) \ \frac{F}{\mu} &=& \^\psi \ + \ \mu \ r \, u  \ , \label{eq:xi1D} \\
\frac{\pa \^\psi}{\pa t}  \ - \ \( \frac{\pa \^\psi}{\pa u} \) \ \frac{F}{\mu} \ + \ \b \, \^\psi &=& \( \frac{\pa F}{\pa x} \) \ \[ \^\phi \ + \ r \, x \] \ - \ \b \, \mu \, r \, u \ . \label{eq:eta1D} \end{eqnarray}

Differentiating \eqref{eq:xi1D} w.r.t. $x$, we get
$$ F'(x) \ \^\phi_u \ = \ 0 \ . $$
This shows that we should distinguish between $F'(x) = 0$ and $F'(x) \not= 0$; the following discussion will show that a similar distinction should be made depending of $F'' (x) = 0$ or $F'' (x) \not= 0$. So we will consider these cases separately.

\subsubsection{Case A: $F' (x) \not= 0$, $F'' (x) \not= 0$}

If $F' (x) \not= 0$, we necessarily have $\pa \^\phi / \pa u = 0$, i.e.
\beq \^\phi (u,t) \ = \ \Phi (t) \ . \eeq
The equation \eqref{eq:xi1D} reads then
\beq \frac{d \Phi}{dt } \ = \ \^\psi (u,t) \ + \ \mu \, r \, u \ , \eeq
and differentiating w.r.t. $u$ we get
$\^\psi_u  =  - \mu r $, which entails
\beq \^\psi (u,t) \ = \ \Psi (t) \ - \ \mu \, r \, u \ . \eeq
Now \eqref{eq:xi1D} reduces to
\beq \Psi (t) \ = \ \Phi' (t) \ . \eeq

We pass to consider \eqref{eq:eta1D}, which reads now
\beq \frac{d^2 \Phi}{d t^2} \ + \ \b \, \frac{d \Phi}{d t} \ = \ \( \Phi \, - \, r \, x \) \ F' (x) \ - \ r \, F (x) \ . \eeq
Differentiating w.r.t. $x$, we get
\beql{eq:f2n0} \[ r x + \Phi (t) \] \ F'' (x) \ = \ 0 \ . \eeq
As we assumed $F'' (x) \not= 0$, we must have (recalling also the relation between $\Phi$ and $\Psi$)
\beq  r = 0 \ , \ \ \ \Phi (t) = 0 \ , \ \ \Psi (t) = 0 . \eeq
In other words, in this case we have no (Lie point) symmetries but the one associated to the $\rho$ constant,
\beq X_0 \ = \ z \ \pa_z \ . \eeq

As anticipated, this is actually a ``ghost'' symmetry, as the variable $z$ does not appear in the equation under study, and is only present to fit in our general formalism developed in previous work \gcite{GGPR,GS17,GW18}. Thus $X_0$, which is always formally present, should be disregarded. We should only look at ``real'' -- as opposed to ``ghost'' -- simple symmetries.

We conclude that:

\medskip\noindent
{\bf Lemma \ref{sec:1Dsymm}.2.} {\it For $F'' (x) \not= 0$, the equation \eqref{eq:1Deq} has no real simple symmetry}.

\subsubsection{Case B: $F' (x) \not= 0$, $F'' (x) = 0$}

We will next consider the case with $F'' (x) = 0$ but $F' (x) \not= 0$. This corresponds to
\beq F(x) \ = \ a \ x \ + \ b \ , \ \ \ \ a \not= 0 \ . \eeq
Note that with the simple change of variable $x = \wt{x} - b/a$ we can always reduce to consider the case with $b=0$. 

The discussion of \eqref{eq:xi1D} is as above, but  now the equation \eqref{eq:eta1D} reads
\beq \frac{d^2 \Phi}{d t^2} \ + \ \b \, \frac{d \Phi}{d t} \ = \ a \, \Phi \ - \ b \, r \ . \eeq
This is immediately solved, yielding
\begin{eqnarray} \Phi (t) &=& \frac{b \, r}{a} \ + \ c_+ e^{- \kappa_+ t} \ + \ c_- e^{- \kappa_- t} \ , \\
\kappa_\pm &: = & \frac12 \( \b \ \pm \ \sqrt{4 a + \b^2} \) \ . \end{eqnarray}

We thus have (using also \eqref{eq:uN} in order to go back to the original variables)
\begin{eqnarray}
\xi &=& r \, \[ (b/a)  \ + \ x \] \ + \ c_+ \, e^{- \kappa_+ t } \ + \ c_- \, e^{- \kappa_- t } \ , \\
\eta &=& r \, v \ - \ c_+ \, \kappa_+ \, e^{- \kappa_+ t} \ - \ c_- \, \kappa_- \, e^{- \kappa_- t} \ . \end{eqnarray}

In other words we have (apart from the ghost symmetry $X_0$) two independent simple symmetries:
\begin{eqnarray}
X_+ &=& e^{- \kappa_+ t} \ \( \pa_x \ - \ \kappa_+ \, \pa_v \) \ , \nonumber \\
X_- &=& e^{- \kappa_- t} \ \( \pa_x \ - \ \kappa_- \, \pa_v \) \ ; \label{eq:symm1Dlin} \end{eqnarray}
and  a W-symmetry given by
\beql{eq:symm1DlinW} Y \ = \ \[ (b/a)  \ + \ x \] \, \pa_x \ + \ v\, \pa_v \ + \ w \, \pa_w \ . \eeq
Note that for $b=0$ (which can always be assumed upon a simple change of variables, as noted above) this is just the generator of the scaling symmetry $x \to \a x$, $v \to \a v$, $w \to \a w$. Thus $Y$ is a pseudo-scaling vector field.

We easily check that
\begin{eqnarray} \[ X_+ , X_- \] &=& 0 \nonumber \\
\[ X_\pm , Y \] &=& X_\pm \ . \end{eqnarray}
We summarize our discussion as:

\medskip\noindent
{\bf Lemma \ref{sec:1Dsymm}.3.} {\it For $F (x) = a x + b$, with $a \not= 0$, the symmetry algebra $\G$ of the equation \eqref{eq:1Deq} has the structure $\G = \X \oplus \Y$, where $\X$ is the Abelian subalgebra spanned by the two real simple symmetries given by \eqref{eq:symm1Dlin}, and $\Y$ is the one-dimensional algebra of W-symmetries generated by the pseudo-scaling vector field  \eqref{eq:symm1DlinW}; the subalgebra $\X$ is an Abelian ideal in $\G$.}

\subsubsection{Case C: $F' (x) = 0$}

In the case $F' (x) = 0$, i.e. $F (x) = c$, it is convenient to operate the simple change of variables \beql{eq:CVFc0} y \ = \ v - c/\b \eeq so to have $c=0$; we will just discuss this case.

For $F(x) = 0$, the equation \eqref{eq:xi1D} yields directly
\beq \^\psi (u,t) \ = \ \^\phi (u,t) \ - \ \mu \, r \, u \ ; \eeq
the equation \eqref{eq:eta1D} reads then
\beq \frac{d \^\phi}{d t^2} \ + \ \beta \ \frac{d \^\phi}{d t} \ = \ 0 \ , \eeq
and thus yields
\beq \^\phi (u,t) \ = \ A (u) \ - \ \frac{e^{- \b t}}{\b} B(u) \ , \eeq
with $A$ and $B$ arbitrary functions. We obtain
\begin{eqnarray*}
\xi &=& r \, x \ + \ A (u) \ - \ \frac{e^{- \b t}}{\b} \, B(u) \ , \\
\eta &=& e^{- \b t} \, B (u) \ + \ \mu \, r \, [ w - u - (\b / \mu ) x] \ = \ e^{- \b t} \, B (u) \ + \ r \, v \ . \end{eqnarray*}

We thus have, beside the ghost symmetry $X_0$, the scaling W-symmetry
\beql{eq:SF0} S \ = \ x \, \pa_x \ + \ v \, \pa_v \ + \ w \, \pa_w \ , \eeq
and two infinite-dimensional sets $\X$ and $\Y$ given respectively by vector fields of the form
\begin{eqnarray}
X_A &=& A(u) \ \pa_x \ , \label{eq:XF0}   \\
Y_B &=& B(u) \ e^{- \b t} \ \( \pa_x \ - \ \b \, \pa_v \) \ . \label{eq:YF0} \end{eqnarray}
(Considering the case $c \not= 0$ would produce the same results, but now the arbitrary functions would depend on $\chi = u + (c /\mu) t$.)

By standard computation we obtain that
\begin{eqnarray*}
\[X_G , X_H \] &=& X_P \ \ \ [P = (\b/\mu) ( G' H - G H' ) ] \ , \\
\[Y_G , Y_H \] &=& 0 \ ; \\
\[X_G , Y_H \] &=& Y_Q \ \ \ [Q = - (\b / \mu) G H' ) \\
\[ X_G , S \] &=& X_P \ \ \ [ P = G + (u/\mu) G' ] \\
\[ Y_G , S \] &=& Y_Q \ \ \ [ Q = G + (u/\mu) G' ] \ . \end{eqnarray*}

We summarize our findings as follows.

\medskip\noindent
{\bf Lemma \ref{sec:1Dsymm}.4.} {\it For $F(x) = 0$, the algebra of real symmetries of equation \eqref{eq:1Deq} is given by $\G = \X \oplus \Y \oplus \S$, where $\X$, $\Y$ are the infinite Lie algebras generated by vector fields \eqref{eq:XF0} and \eqref{eq:YF0} respectively, having also the structure of Lie module over the invariants, and $\S$ is the one-dimensional algebra generated by \eqref{eq:SF0}. The subalgebra $\X \oplus Y$ is an ideal in $\G$, and the subalgebra $\Y$ is an Abelian ideal in $\G$.}
\bigskip

%\newpage

\section{The $n$-dimensional isotropic case.}
\label{sec:n_dim}

We will now consider the isotropic case; we recall this is meant in the sense discussed in Remark 14. That is, we have no hypothesis (yet) on $F(\xb)$, but we assume
\beql{eq:isotropic} \b_{(1)} \ = \ \b_{(2)} \ = \ ... \ = \ \b_{(n)} \ = \ \b \ ; \ \ \
\mu_{(1)} \ = \ \mu_{(2)} \ = \ ... \ = \ \mu_{(n)} \ = \ \mu \ . \eeq

This makes little difference in the general case for the invariants, but simplifies the discussion of symmetries.

\medskip\noindent
{\bf Remark 16.} As for invariants, our general discussion of Sect.\ref{sec:gen_inv} holds true, and we will have invariants only in the case $\Fb = {\bf c}$; the only difference is that in the case with constant force, the form of the invariants get simplified, i.e. we have still \eqref{eq:chirho}, but now with $\rho^i = c^i/ \mu$. \EOR
\bigskip

As for symmetries, in fact, with the assumption \eqref{eq:isotropic}
the coefficients of the would-be symmetry vector fields are
\begin{eqnarray}
\xi^i &=& \^\phi^i (\ub,t) \ + \ \sum_{j=1}^n  \^R^{i}_{\ j} \ x^j \ , \label{eq:xiiso} \\
\eta^i &=& \^\psi^i (\ub,t) \ + \ \sum_{j=1}^n \^R^{i}_{\ j} \ \( \mu \, w^j \ - \ \b \, x^j \) \ ; \label{eq:etaiso} \end{eqnarray}
the odd-numbered equations  $\E_{2i-1}$ are
\begin{eqnarray} \frac{\pa \^\phi^i}{\pa t} \ - \ \frac{1}{\mu} \sum_{j=1}^n \( \frac{\pa \^\phi^i}{\pa u^j} \) \ F^j &=& \^\psi^i (t) \ + \ \mu \ \sum_{j=1}^n \, \(  \^R^{i}_{\ j} \, u^j \) , \label{eq:redxiiso} \end{eqnarray}
while the even-numbered equations $\E_{2i}$ read
\begin{eqnarray} \frac{\pa \^\psi^i}{\pa t} \ - \ \frac{1}{\mu} \sum_{j=1}^n \( \frac{\pa \^\psi^i}{\pa u^j} \) \ F^j \ + \ \b \, \^\psi^i &=& \sum_{j=1}^n \( \frac{\pa F^i}{\pa x^j} \) \ \[ \^\phi^j \ + \ \sum_{k=1}^n \^R^{j}_{\ k} \, x^k \] \nonumber \\
 & & \ - \ \b \, \mu \sum_{j=1}^n \( \^R^{i}_{\ j} \, u^j \) \ . \label{eq:redetaiso} \end{eqnarray}

Differentiating \eqref{eq:redxiiso} w.r.t. $x^k$ we get
\beql{eq:phihatFiso} \sum_{j=1}^n \( \frac{\pa \^\phi^i}{\pa u^j} \) \ \( \frac{\pa F^j}{\pa x^k} \) \ = \ 0 \ . \eeq
The consequences of this relation depend on the detailed form of $F$ (e.g., for $F=c$ this is identically satisfied). In fact, we will generalize the discussion conducted in the one-dimensional case, and proceed accordingly.

\subsection{Regular and fully regular force field}
\label{sec:regular}

Our discussion above suggests the following definition.

\medskip\noindent
{\bf Definition 1.} {\it If the matrix $(D_x F)$ with entries
$(D_x F)^j_{\ k} = \( \pa F^j / \pa x^k \)$ is non-singular, we say that the force field ${\bf F} (\xb)$ is \emph{regular}.}

\medskip\noindent
{\bf Remark 17.} The case $F=c$ is not regular; in the linear case $F^i (x) = L^i_{\ j} x^j + K^i$, the regularity of $F$ corresponds to the regularity of the matrix $L$. \EOR
\bigskip

For $F$ regular, \eqref{eq:phihatFiso} implies that the matrix
\beq (D_u \^\phi) \ := \ \( \pa \^\phi^i / \pa u^j \) \eeq is identically zero, which in turn implies
\beq \^\phi^i (\ub , t) \ = \ \Phi^i (t) \ . \eeq

Now the odd-numbered equations  $\E_{2i-1}$ are
\begin{eqnarray} \frac{d \Phi^i}{d t} &=& \^\psi^i (\ub,t) \ + \ \mu \ \sum_{j=1}^n \, \(  \^R^{i}_{\ j} \, u^j \) \ .  \end{eqnarray}
Differentiating these w.r.t. the $u^j$ we obtain
\beq \^\psi (\ub,t) \ = \ \Psi (t) \ - \ \mu \ \sum_j R^i_{\ j} \, u^j \ , \eeq
and the equations $\E_{2i-1}$ reduce now to
\beql{eq:eqoddSOFR} \Psi^i (t) \ = \ \frac{d \Phi^i}{d t} \ . \eeq

Let us pass to consider the even-numbered equations $\E_{2i}$. These now read
\beql{eq:eqevenSOFR} \frac{d^2 \Phi^i}{d t^2} \ + \ \b \, \frac{d \Phi^i}{d t} \ + \ \sum_j \^R^i_{\ j} \, F^j \ = \ \sum_j \frac{\pa F^i}{\pa x^j} \ \[ \Phi^j \ + \ \sum_k \^R^j_{\ k} \, x^k \] \ . \eeq
Considering the second derivative of these equations, once w.r.t. $t$ and once w.r.t. $x^j$, we get
\beql{eq:Freg2} \sum_k \ \frac{\pa^2 F^i}{\pa x^j \pa x^k} \ \frac{d \Phi^k}{d t} \ = \ 0 \ . \eeq

We define symmetric matrices $\mathcal{F}^{(i)}$ with entries
\beql{eq:Freg2mat} \( \mathcal{F}^{(i)} \)_{kj} \ = \ \frac{\pa^2 F^i}{\pa x^k \pa x^j} \ , \eeq so that (with $\Phi$ the vector of components $\Phi^i$)  the relation \eqref{eq:Freg2} reads
\beql{eq:SORphi}  \mathcal{F}^{(i)}  \ \Phi \ = \ 0 \ . \eeq
This suggests a new definition.

\medskip\noindent
{\bf Definition 2.} {\it Consider a vector force field ${\bf F} (\xb)$ in ${\bf R}^n$ and its associated matrices $\( \mathcal{F}^{(i)} \)$ defined by \eqref{eq:Freg2mat}; if all the $n$ matrices $\( \mathcal{F}^{(i)} \)$ are regular, we say  we say that ${\bf F} (\xb)$ is \emph{second order fully regular}. If at least one of the matrices $\( \mathcal{F}^{(i)} \)$ is regular, we say that ${\bf F} (\xb)$ is \emph{second order regular}.}

\medskip\noindent
{\bf Remark 18.} Note that if $F$ is second order regular, then it is necessarily also regular. On the other hand, linear vector fields provide a simple example of $F$ which is regular but not second order regular. \EOR

\medskip\noindent
{\bf Remark 19.} Consider $F$ of the functional form
$$ F^i (\xb) \ = \ h(|\xb|) \ x^i \ , $$ with $h$ a nonzero function. In this case we have an isotropic force field, which actually descends from an isotropic potential $H(|\xb|)$. It is easily seen that $F$ is regular, and that $F$ is also second order regular unless $h$ is constant, i.e. unless $F$ is linear. This, albeit a special case in mathematical terms, is of course a relevant one in physical ones; it provides justification for focusing on the regular or second order regular case. \EOR

\subsection{Symmetries}

We are now ready to give a full description of symmetries for force fields which are second order regular or at least regular.

\subsubsection{Case A: second order regular $F$}

As already remarked, the relation \eqref{eq:Freg2} with our notation reads \eqref{eq:SORphi}. If $F$ is second order fully regular, we act on this by $(\mathcal{F}^{(i)})^{-1}$ (for any $i = 1,...,n$) and obtain $d \Phi^k / d t= 0$ for all $k = 1,...,n$. That is, we have
\beql{eq:phiSOFR} \Phi^i (t) \ = \ c^i \eeq with $K^i$ constants. The equations $\E_{2i}$. now read
\beq \sum_j \( \^R^i_{\ j} \, F^j \ - \ \frac{\pa F^i}{\pa x^j} \ \[ c^j \ + \ \sum_k \^R^j_{\ k} \, x^k \] \) \ = \ 0 \ . \eeq
Using the notation introduced above for $(D_x F)$, this is rewritten vector notation as
\beql{eq:RFSOR} \^R \, \Fb \ = \ (D_x F) \, \( {\bf c} \ + \ \^R \, \xb \) \ ; \eeq
acting on the left by $M := (D_x F)^{-1} $, we get
\beq {\bf c} \ = \ M \, \^R \, \Fb \ - \ \^R \, \xb \ . \eeq

This shows immediately that $\^R = 0$ implies ${\bf c} = 0$, hence $\Phi^i (t) = 0$ and from \eqref{eq:eqoddSOFR} also $\Psi^i = 0$; note \eqref{eq:xiiso} and \eqref{eq:etaiso} imply then $\xi^i = 0 = \eta^i$. In other words, no symmetries with $\^R = 0$ are possible.

We summarize our general finding as follows:

\medskip\noindent
{\bf Lemma \ref{sec:n_dim}.1.} {\it In the isotropic case and for $F$ second order fully regular, the equation \eqref{eq:BM} has no regular (deterministic or random) simple symmetry, while it may have real simple W-symmetries}.

\medskip\noindent
{\bf Remark 20.} The possibility of having W-symmetries, i.e. with $\^R \not= 0$, should be analyzed in detail depending on $\Fb$, and it appears no general statements can be given without solving the equations for the case at hand. Similarly, if $F$ is second order regular but not fully regular, the reasoning leading to \eqref{eq:phiSOFR} only applies to the $\mathcal{F}^{(i)}$ which are regular, and hence \eqref{eq:phiSOFR} only holds for the corresponding $\Phi^i$, and some detailed discussion would be needed. Again, no general statements can be given. \EOR

\medskip\noindent
{\bf Remark 21.} We note that if $\Fb$ admits higher order terms $\Fb_{(n)}$ (i.e. it is a finite order polynomial, $\Fb_{(n)}$ being the highest order terms), then \eqref{eq:RFSOR} implies, with an obvious notation,
\beql{eq:RFSORn} \^R \, \Fb_{(n)} \ = \ (D_x F_{(n)}) \, \^R \, \xb  \ , \eeq
or equivalently $M_{(n)} \^R \Fb_{(n)} = \^R \xb$; this may be a good starting point for the concrete analysis of a given $\Fb$. \EOR

\subsubsection{Case B: linear regular $F$}

We consider next the case where $F$ is regular but not second order regular; that is, $F$ is linear and regular; thus
$$ F^i (x) \ = \ L^i_{\ j} \, x^j \ + \ K^i $$ with $L$ a regular matrix.
Then \eqref{eq:eqoddSOFR} holds, and \eqref{eq:eqevenSOFR} reads
\beql{eq:eqevenggg}  \frac{d^2 \Phi^i}{d t^2} \ + \ \b \, \frac{d \Phi^i}{d t} \ = \ L^i_{\ j} \ \Phi^j \ + \ \( L^i_{\ j} \, \^R^j_{\ k} \ - \ \^R^i_{\ j} \, L^j_{\ k} \) \, x^k \ - \ \^R^i_{\ j} \, K^j \ . \eeq

Differentiating these w.r.t. $x^k$, we obtain that $\^R$ must commute with $L$,
$$ \[ L \, , \, \^R \] \ = \ 0 \ . $$
This requirement can be more or less severe depending on the form of $L$. 

\medskip\noindent
{\bf Remark 22.} Note that as $L$ is regular, we can eliminate the constant terms by a simple shift in $\xb$, $\xb = \wt{\xb} - {\bf s}$, with ${\bf s} = - L^{-1} {\bf K}$ (note we leave the $\vb$ and $\wb$ variables unchanged). In these variables we have 
$$ d \wt{x}^i \ = \ v^i dt \ , \ \ d v^i \ = \ \( L^i_{\ j} x^j - \b v^i \) dt + \mu d w^i \ . $$ 
Moreover, once we have a purely linear force (no constant terms), this can be further simplified by a new change of variables, $\wt{\xb} = M^i_{\ j} \^x^j $, where $M$ is the diagonalizing matrix\footnote{Note that in principles $L$ could have a nilpotent part and hence its normal form could contain some Jordan blocks. We are however primarily interested in the case the force field originates from a potential: in this case $L$ is semisimple and its Jordan form is just diagonal.} for $L$, $M^{-1} L M = \mathrm{diag} ( \la_1 , ... , \la_n )$. This should be accompanied by operating the same change of variables on the $v^i$ and the $w^i$ variables, $v^i = M^i_{\ j} \^v^j$, $w^i = M^i_{\ j} \^w^j$. In the following we will thus operate directly with $L$ diagonal,
$$ L \ = \ \mathrm{diag} ( \la_1 , ... , \la_n ) \ . $$
This will also help comparison with the scalar case. \EOR

\bigskip\noindent
{\it A. Regular symmetries}
\medskip

We will first focus on the case $\^R = 0$, i.e. on regular symmetries; now \eqref{eq:eqevenSOFR} is just
\beql{eq:Phi_iso_lin} \frac{d^2 \Phi^i}{d t^2} \ + \ \b \, \frac{d \Phi^i}{d t} \ = \ L^i_{\ j} \ \Phi^j  \ . \eeq

With our assumption of $L$ diagonal, the system consists of $n$ decoupled equations, each in the form
$$ \frac{d^2 \Phi^i}{d t^2} \ + \ \b \, \frac{d \Phi^i}{d t} \ = \ \lambda_{(i)} \ \Phi^i  \ . $$
Setting now the shorthand notation
\beql{eq:kappaenne} \kappa^{(i)}_\pm \ := \ \frac{1}{2} \ \[ \b \ \pm \ \sqrt{\b^2 \, + \, 4 \, \lambda_{(i)} } \] \ , \eeq
we have solutions (with $c^{(i)}_\pm$ arbitrary constants)
$$ \Phi^i (t) \ = \ c^{(i)}_+ \ \exp[ - \kappa_+^{(i)} t ] \ + \ c^{(i)}_- \ \exp[ - \kappa_-^{(i)} t ] \ . $$
It follows immediately
$$ \Psi^i (t) \ = \ \frac{d \Phi^i}{d t} \ = \ - \ \(  c^{(i)}_+ \, \kappa_+^{(i)} \ \exp[ - \kappa_+^{(i)} t ] \ + \ c^{(i)}_- \, \kappa_-^{(i)} \ \exp[ - \kappa_-^{(i)} t ] \) \ . $$

We thus have $2n$ symmetries, which are read from the above by setting only one of the $2n$ constants $c^{(i)}_\pm$ different from zero. Note that $\kappa_\pm$ are constants; thus it is immediate to see that all these symmetries commute with each other. (Nothing changes in this respect if the system with diagonal $L$ was obtained upon changes of coordinates, as all of these only involve constant factors.)  

It is also immediate to see these are, according to our classification, regular \emph{deterministic} symmetries; our discussion shows that there are no regular random symmetries.

Recalling that we are assuming $\^R = 0$, hence $\^\phi$ and $\^\psi$ give directly the coefficients $\xi$ and $\eta$, see \eqref{eq:redxifin} and \eqref{eq:redetafin}, the Abelian Lie algebra of symmetries is spanned by the vector fields ($i = 1,...,n$)
\beql{eq:symmlinFiso0} X_+^{(i)} \ = \ e^{- \kappa_+ t} \( \frac{\pa}{\pa x^i} \ - \ \kappa_+ \, \frac{\pa}{\pa v^i} \) \ , \ \ \ X_-^{(i)} \ = \ e^{- \kappa_- t} \( \frac{\pa}{\pa x^i} \ - \ \kappa_- \, \frac{\pa}{\pa v^i} \) \ . \eeq

We summarize our discussion concerning regular (i.e. deterministic or random) symmetries as follows:

\medskip\noindent
{\bf Lemma \ref{sec:n_dim}.2.} {\it In the $n$-dimensional isotropic case and for $F$ regular but not second order regular, the equation \eqref{eq:BM} has $2 n$ real regular simple symmetries, given in \eqref{eq:symmlinFiso0}; these are deterministic symmetries and span an Abelian Lie algebra.}

\bigskip\noindent
{\it B. W-symmetries}
\medskip

In order to discuss W-symmetries as well, let us go back to eq.\eqref{eq:eqevenggg}. Differentiating \eqref{eq:eqevenggg} w.r.t. $x^k$ we immediately obtain $L^i_{\ j } \^R^j_{\ k} - \^R^i_{\ j} L^j_{\ k} = 0$, i.e. $L$ and $\^R$ must commute,
\beql{eq:eqevengggx} \[ L \, , \, \^R \] \ = \ 0 \ . \eeq
With our assumption that $L$ is diagonal, this condition results simply in
\beql{eq:eqevengggx2} \( \la_{(i)} \ - \ \la_{(j)} \) \ \^R^i_{\ j} \ = \ 0 \ . \eeq
Thus we can have nonzero $\^R^i_{\ j}$ only if $\la_{(i)} \not= \la_{(j)}$.

In other words, W-symmetries will depend on degeneracies among eigenvalues of $L$.
One should thus discuss all possible subcases of partial degeneration; we will limit to consider the cases of total degeneration ($\la_{(i)} = \la$ for all $i=1,...,n$) and of total non-degeneration ($\la_{(i)} \not= \la_{(j)}$ for $i \not= j$).

\bigskip\noindent
{\it B.1. W-symmetries, fully degenerate $L$}
\medskip

In the fully degenerate case, $\la_{(1)} = .... = \la_{(n)} = \la \not= 0$. Then $ L = \la I$ and the equations \eqref{eq:eqevengggx} are automatically satisfied; that is, there is no restriction on $\^R$. 

The equations are now 
\beq \frac{d^2 \Phi^i}{d t^2} \ + \ \b \, \frac{d \Phi^i}{d t} \ - \ \la \ \Phi^i \ = \ 0 \ . \eeq
With the notation
$$ \ga \ := \ \sqrt{\b^2 + 4 \la} \ , $$ their solution is, with $c_\pm^i$ arbitrary constants, 
\beq \Phi^i (t) \ = \ c_+^i \ \frac{\exp[- (t/2) (\b + \ga )]}{\b + \ga } \ + \ c_-^i  \ \frac{\exp[(t/2) (\b + \ga )]}{\b - \ga } \ ; \eeq
this implies of course
\beq \Psi^i (t) \ = \ \frac{d \Phi^i}{d t} \ = \ - \frac{c_+^i}{2} (\b + \ga)  \ \frac{\exp[- (t/2) (\b + \ga )]}{\b + \ga } \ + \ \frac{c_-^i}{2} (\b + \ga)  \ \frac{\exp[(t/2) (\b + \ga )]}{\b - \ga } \ . \eeq
Going back to \eqref{eq:redxifin}, \eqref{eq:redetafin} -- and recalling again \eqref{eq:uN} -- we have vector fields as in \eqref{eq:symmlinFiso0}, and moreover vector fields associated to the elements of the $\^R$ matrix (those we are looking for at present) which are
\beq X_R \ = \ \^R^i_{\ j} \, x^j \, \frac{\pa}{\pa x^i}  \ + \ \^R^i_{\ j} \, v^j \, \frac{\pa}{\pa v^i} \ + \ \^R^i_{\ j} \, w^j \, \frac{\pa}{\pa w^i} \ . \eeq
In particular, the vector fields associated to the diagonal part of $R$ are scaling ones (acting in the same way on the $\ub$, the $\vb$ and the $\wb$ variables), i.e. we have symmetries (no sum on the $i$ index here)
\beql{eq:Sinov} S_{(i)} \ = \ x^i \, \frac{\pa}{\pa x^i}  \ + \ v^i \, \frac{\pa}{\pa v^i} \ + \ w^i \, \frac{\pa}{\pa w^i} \ . \eeq
Vector fields associated to the off-diagonal part of $\^R$ correspond to rotations (again acting in the same way on the $\ub$, the $\vb$ and the $\wb$ variables), i.e. -- again with no sum on the repeated $i$ and $j$ indices -- we have symmetry vector fields
\beq R_{(ij)} \ = \ \( x^j \, \frac{\pa}{\pa x^i} \ - \ x^i \, \frac{\pa}{\pa x^j} \) \ + \ \( v^j \, \frac{\pa}{\pa v^i} \ - \ v^i \, \frac{\pa}{\pa v^j} \) \ + \ \( w^j \, \frac{\pa}{\pa w^i} \ - \ w^i \, \frac{\pa}{\pa w^j} \) \ . \eeq Needless to say, the ensemble of the vector fields $X_R$, i.e. the $S_{(i)}$ and the $R_{(i,j)}$ span the algebra of the linear conformal group. 

Note that independent scalings in different directions are possible, as also implied by our equation \eqref{eq:Rform}; see also \eqref{eq:LCG}. This is also obvious from the form of our Ito equations, which with the present assumptions just read $$ d x^i = v^i dt \ , \ \ d v^i = (\la x^i - \b v^i ) dt + \mu dw^i \ . $$

Let us now consider the commutation relations among symmetry generators in \eqref{eq:symmlinFiso0} and those for the W-symmetries. By standard computations, we get 
\begin{eqnarray*}
\[ X^{(i)}_\pm \ , \ S_{(j)} \] &=& \de^i_j \ X^{(i)}_\pm \ , \\
\[ X^{(i)}_\pm \ , \ R_{(jk)} \] &=& - \, \de^i_j \, X^{(k)}_\pm \ + \ \de^i_k \, X^{(j)} \ . \end{eqnarray*}

\bigskip\noindent
{\it B.2. W-symmetries, fully non-degenerate $L$}
\medskip

In the fully non-degenerate case, only diagonal terms in $\^R$ can be nonzero. Apart from this, the computations are exactly the same as in point {\it B.2} above. 

That is, we have vector fields as in \eqref{eq:symmlinFiso0}, corresponding to regular symmetries, and moreover vector fields associated to the diagonal elements $r_{(i)}$ of the $\^R$ matrix, i.e. the scaling vector fields $S_{(i)}$ defined in \eqref{eq:Sinov}. Note that now the Ito equations are \beql{eq:Fnlindiag} d x^i = v^i dt \ , \ \ d v^i = (\la_{(i)} x^i - \b v^i ) dt + \mu dw^i \ ; \eeq the differences between the different $\la_{(i)}$ cause rotations in the $\xb$, $\vb$ and $\wb$ spaces not to be a symmetry.
\bigskip

We summarize the situation in the following

\medskip\noindent
{\bf Lemma \ref{sec:n_dim}.3.} {\it In the $n$-dimensional isotropic case and for $F$ regular but not second order regular, the equation \eqref{eq:BM} has a non-Abelian symmetry algebra $\G = \X \oplus \Y$, where $\X$ is the Abelian subalgebra of regular symmetries described in Lemma \ref{sec:n_dim}.2 and generated by the vector fields \eqref{eq:symmlinFiso0}; while $\Y$ is the subalgebra of W-symmetries. In the fully degenerate case, $\Y$ is the algebra of the Linear Conformal Group, while in the fully non-degenerate case, the subalgebra $\Y$ is also Abelian and is spanned by the scaling vector fields \eqref{eq:Sinov}.}

\subsubsection{Case C: constant $F$}

In the case $$ F^i (\xb) \ = \ c^i \ , $$ we can always reduce to consider
the free particle through (the vector version of) the simple change of variables \eqref{eq:CVFc0}, i.e. 
through $v^i = y^i + c^i/\b$; this amounts to passing to a moving frame. We will thus just consider this case, and set from now on  
$$ c^i \ = \ 0 \ . $$ The reduced equations \eqref{eq:redxiiso}, \eqref{eq:redetaiso} read
\begin{eqnarray}
\frac{\pa \^\phi^i}{\pa t} &=& \^\psi^i \ + \ \mu \ \sum_{j=1}^n \^R^i_{\ j} \, u^j \label{eq:rxiic0} \\
\frac{\pa \^\psi^i}{\pa t} & = & - \b \, \[ \^\psi^i \ + \ \mu \ \sum_{j=1}^n \^R^i_{\ j} \, u^j \] \ . \label{eq:retaic0} \end{eqnarray}
Obviously, equation \eqref{eq:rxiic0} immediately yields
\beq \^\psi^i \ = \ \frac{\pa \^\phi^i}{\pa t} \ - \ \mu \ \sum_{j=1}^n \^R^i_{\ j} \, u^j \ ; \eeq we can then substitute in \eqref{eq:retaic0} according to this. We obtain simply
\beq \frac{\pa^2 \^\phi^i}{\pa t^2} \ + \ \b \ \frac{\pa \^\phi^i}{\pa t} \ = \ 0 \ , \eeq
which yields at once 
\begin{eqnarray*}
\^\phi^i (t,\ub) &=& A^i (\ub) \ - \ \frac{e^{- \b t}}{\b} \ B^i (\ub ) \ , \\
\^\psi^i (t,\ub) &=& e^{- \b t} \ B^i (t,\ub) \ - \ \mu \^R^i_{\ j} \ u^j \ , \end{eqnarray*}
where the $A^i$ and $B^i$ are arbitrary functions.

Finally, recalling \eqref{eq:xiiso}, \eqref{eq:etaiso} and \eqref{eq:uN}, we have 
\begin{eqnarray}
\xi^i &=& A^i (\ub) \ - \ \frac{^{- \b t}}{\b} \ B^i (\ub) \ + \ \^R^i_{\ j} \ x^j \ , \\
\eta^i &=& e^{- \b t} \ B^i (\ub) \ + \ \^R^i_{\ j} \ v^j \ . \end{eqnarray}

We conclude that we have two families of random symmetries depending on arbitrary functions of $\{ u^1 , ... , u^n \}$, i.e.\begin{eqnarray}
X^i_\varrho &=& \varrho (\ub) \ \frac{\pa}{\pa x^i} \ , \label{eq:XnisoC} \\
Y^i_\ga &=& \ga (\ub ) \ e^{- \b t} \[ \frac{\pa}{\pa x^i} \ - \ \b  \, \frac{\pa}{\pa v^i} \] \ ; \label{eq:YnisoC} \end{eqnarray}
this shows that we have an infinite dimensional Lie (sub)algebra of regular symmetries $\G_0 = \X \oplus \Y$, which is two-dimensional as a Lie module over the ring of invariants, generated itself by $\{ u^1,...,u^n \}$. Note that when we go back to the original variables, in which $F^i (\xb ) = c^i$, the $u^i$ are replaced by the invariants $\chi^i$ met in our discussion above, see \eqref{eq:chirho}.

Let us now consider the commutation relations among these vector fields. By elementary computations we obtain
\begin{eqnarray*}
\[ X_g^i \ , \ X_h^j \] &=& \frac{\beta}{\mu} \ h \ \( \frac{\pa g}{\pa u^j} \) \, \frac{\pa }{\pa x^i} \ - \ \frac{\beta}{\mu} \ g \ \( \frac{\pa h}{\pa u^i} \) \, \frac{\pa }{\pa x^j} \ := \ X_G^i \ - \ X_H^j \ ; \\
\[ Y_g^i \ , \ Y_h^j \] &=& 0 \ ; \\
\[ X_g^i \ , \ Y_h^j \] &=& - \, \frac{\b}{\mu} \, e^{- \b t} \, g \, \( \frac{\pa h}{\pa u^i} \) \, \[ \frac{\pa}{\pa x^j} \ - \ \b \,  \frac{\pa}{\pa v^j} \] \ := \ Y_H^j \ . \end{eqnarray*}

We summarize our discussion as follows:

\medskip\noindent
{\bf Lemma \ref{sec:n_dim}.4.} {\it In the $n$-dimensional isotropic case and for $F^i (\xb) = 0$, the equation \eqref{eq:BM} has an infinite dimensional Lie algebra of regular symmetries $\G_0$, spanned by vector fields of the form \eqref{eq:XnisoC}, \eqref{eq:YnisoC}. The set $\G$ is $2 n$-dimensional as a Lie module over the ring of invariants, the latter being generated by $\{ \chi^1 , ... , \chi^n \}$ defined in \eqref{eq:chirho}. The fields defined in \eqref{eq:XnisoC} and in \eqref{eq:YnisoC} form two subalgebras $\X$ and $\Y$ in $\G_0 = \X \oplus \Y$; each of them is also a Lie module. Moreover, $\Y$ is Abelian and an Abelian ideal in $\G_0$.}

\medskip\noindent
{\bf Remark 23.} The results obtained for constant external force in the isotropic case are essentially equivalent, modulo the slightly different form of the invariants $\chi^i$ (which replace the $u^i$), to those obtained in the case ${\bf F} = {\bf 0}$. \EOR

\medskip\noindent
{\bf Remark 24.} Finally, we stress that in addition to $\G_0$ we have the W-symmetries associated to the elements of $\^R$, i.e. (no sum on repeated indices)
\beq Z^i_{\ j} \ = \ \^R^i_{\ j} \[ x^j \, \frac{\pa}{\pa x^i} \ + \  v^j \, \frac{\pa}{\pa v^i} \ + \ w^j \, \frac{\pa}{\pa w^i} \] \ . \eeq
In particular, the symmetries associated to the diagonal part of $\^R$ are scaling ones, while those associated to the off-diagonal part do - recalling that $\^R$ is a generator of the conformal linear group in $n$ dimensions -- generate the group $\mathrm{SO}(n)$ of simultaneous rotations in the $x$, $v$ and $w$ spaces. \EOR

\section{Symmetry integration of the Ornstein-Uhlenbeck process}
\label{sec:integration}

As recalled in Sect.\ref{sec:symmstoch}, the main motivation to study symmetries of a stochastic equations (and to restrict to simple symmetries) lies in that symmetries allow to integrate, or at least reduce, the stochastic equation.

In our case, the discussion of Sect.\ref{sec:n_dim} shows that -- at least in the isotropic case -- there are symmetries only for $F$ constant or linear, so we have only to analyze these cases. Actually, in both these cases we have -- directly in the constant case, and in general upon a change of coordinates in the linear case, see Remark 22 -- a collection of one-degree-of-freedom problems. We can thus just study this kind of problems. {We will only consider regular symmetries.}

\subsection{Regular linear force}

We will start from \eqref{eq:Fnlindiag}, and actually, for ease of notation, just consider one of the subsystems in which this splits, i.e.
\beql{eq:Flin} \begin{cases} d x^i \ = \ v^i \ dt \ , & \\ d v^i \ = \ [ \la_{(i)} \, x^i \ - \ \b \, v^i ] \ d t \ + \ \mu \ d w^i & . \end{cases} \eeq

The symmetry analysis conducted above, and whose results are summarized in Lemma \ref{sec:n_dim}.2, tells us that for each degree of freedom (i.e. for each fixed $i=1,...,n$) we have two symmetry vector fields, given by \eqref{eq:symmlinFiso0}.

We should then pass to \emph{symmetry-adapted variables}, i.e. to variables $y^i_\pm$ such that $$ X^{(i)}_\pm \ = \ \frac{\pa}{\pa y^i_\pm} \ . $$ It is rather clear this will not need to mix the different indices $i$, i.e. we will have for each $i=1,...,n$ a change of variables \beql{eq:xvylin} (x^i,v^i) \ \to \  (y^i_+,y^i_-) \ . \eeq

This is determined by requiring the variables $y^i_\pm$ to satisfy the system of equations
\beql{eq:Xysym} X^{(i)}_+ (y^i_+)=1 , \ X^{(i)}_+ (y^i_-)=0 \ ; \ \ X^{(i)}_- (y^i_+)=0 , \ X^{(i)}_- (y^i_-)=1 \ . \eeq As the $X^{(i)}_\pm$ are first order differential operators, the equations are promptly solved by the method of characteristics through standard computations. The final outcome is that the general solution to \eqref{eq:Xysym} is provided by
\begin{eqnarray}
y^i_+ &=& \frac{\exp[{\kappa^{(i)}_+ t}]}{\kappa^{(i)}_- \, - \, \kappa^{(i)}_+} \ (\kappa^{(i)}_- \, x \ + \ v ) \ + \ H^{(i)}_+ (t) \ , \nonumber \\
y^i_- &=& \frac{\exp[{\kappa^{(i)}_-  t}]}{\kappa^{(i)}_+ \, - \, \kappa^{(i)}_-} \ (\kappa^{(i)}_+ \, x \ + \ v ) \ + \ H^{(i)}_- (t) \ , \end{eqnarray}
where $H^{(i)}_\pm (t)$ are arbitrary functions, which might be set to zero for the sake of simplicity. It is immediate to check these variables satisfy \eqref{eq:Xysym}; moreover we have $X^{(i)}_\pm ( y^j_\pm ) = 0$ for $i \not= j$.

We should now determine what is the dynamics undergone by the $y^i_\pm$ when $(x^i,v^i)$ evolve according to \eqref{eq:Flin}. For this, it suffices to apply the Ito rule taking into account the special form of the diffusion matrix $\s$.

This gives at first a rather involved formula, which we do not report here, in which the coefficients $\kappa^{(i)}_\pm$ appear. When we substitute for these according to \eqref{eq:kappaenne}, we get
\begin{eqnarray}
d y^i_+ &=& (H^{(i)}_+)' (t) \, dt \ - \ \frac{\exp[(1/2)(\b + \sqrt{\b^2 + 4 \la_{(i)} t})]}{\sqrt{\b^2 \, + \, 4 \la_{(i)} }} \ \mu \ d w^i \nonumber \\
&:=& (H^{(i)}_+)' (t) \, dt \ + \ \a_+^{(i)} (t) \, d w^i \ , \nonumber \\
d y^i_- &=& (H^{(i)}_-)' (t) \, dt \ + \ \frac{\exp[(1/2)(\b - \sqrt{\b^2 + 4 \la_{(i)} t})]}{\sqrt{\b^2 \, + \, 4 \la_{(i)} }} \ \mu \ d w^i \nonumber \\
&:=& (H^{(i)}_-)' (t) \, dt \ + \ \a^{(i)}_- (t) \, d w^i \ . \end{eqnarray}
The solutions are therefore easily expressed in terms of Ito integrals,
\begin{eqnarray}
y^{(i)}_+ (t) \ = \ H^{(i)}_+ (t) \ + \ \int \a^{(i)}_+ (t) \ d w^i (t) \ , \nonumber \\
y^{(i)}_- (t) \ = \ H^{(i)}_- (t) \ + \ \int \a^{(i)}_- (t) \ d w^i (t) \ . \end{eqnarray}
It suffices then to invert the change of variables \eqref{eq:xvylin} to give the solution in terms of the original $(x^i,v^i)$ variables.

\medskip\noindent
{\bf Remark 25.} It may be of some interest to discuss what would happen if one is not passing directly to symmetry adapted variables by looking for solutions to \eqref{eq:Xysym}, but proceeds in a less systematic way.
The form of the vector fields $X^{(i)}_\pm$ suggests to operate a linear change of variables, and pass to consider
\beql{eq:y} y^i_\pm \ = \ A^{(i)}_\pm \, x^i \ + \ v^i \ , \eeq
with $A_\pm = \kappa_\pm$ the constants defined in \eqref{eq:kappaenne}.
For these variables we have, under the dynamics \eqref{eq:Flin} and applying the Ito rule,
\begin{eqnarray*} d y^i_\pm &=& A^{(i)}_\pm \ d x^i \ + \ d v^i \\
&=& \[ \( A^{(i)}_\pm \ - \ \b \) \, y^i_\pm \ + \ \( \la_{(i)} \ - \ \( A^{(i)}_\pm \ - \ \b \) \, A^{(i)}_\pm \) \, x^i \] \ dt \ + \ \mu \, d w^i \ . \end{eqnarray*}
The coefficient of $x^i dt$ in the r.h.s. vanishes for
$A^{(i)}_\pm = (1/2)(\b \pm \sqrt{\b^2 \ + \ 4 \, \la_{(i)} })$, i.e. exactly for \beql{eq:Ak} A^{(i)}_\pm \ = \ \kappa^{(i)}_\pm \ . \eeq
That is, under \eqref{eq:Flin} we have that with the choice \eqref{eq:Ak} the variables defined in \eqref{eq:y} evolve according to
\beql{eq:ysep} \begin{cases} d y^i_+ \ = \ B^{(i)}_+ \, y^i_+ \, dt \ + \ \mu \, d w^i & , \\
d y^i_- \ = \ B^{(i)}_- \, y^i_- \, dt \ + \ \mu \, d w^i & , \end{cases} \eeq
where the constants $B^{(i)}_\pm$ are defined as
\beq B^{(i)}_\pm \ := \  \kappa^{(i)}_\pm \ - \ \b \ . \eeq
We have thus reached a \emph{separation of variables}.

It should be noted that we do \emph{not} have equations in integrable form; this corresponds to the fact we have not taken variables which are \emph{fully} adapted to the vector fields $X^{(i)}_\pm$, as we disregarded the pre-factor $\exp[- \kappa_\pm t]$ in them.\footnote{We could then look for a second change of variables, in the form
$ z^i_\pm = q^{(i)}_\pm \exp[ \a^{(i)}_\pm \ t] y^i_\pm$
with $q^{(i)}_\pm$ constants and with $\a^{(i)}_\pm $ such to get an integrable equation for  $z^i_\pm$. In this way we reach of course the same result as above.} \EOR

\subsection{Constant force}

In the case of constant external force, $F^i = c^i$, we have
\beql{eq:FC} \begin{cases} d x^i \ = \ v^i \ dt & \\ d v^i \ = \ [ c^i \ - \ \b \, v^i ] \ d t \ + \ \mu \ d w^i & . \end{cases} \eeq
We have seen that the symmetry Lie module (over the ring of invariants, generated by $\{ \chi^1,...,\chi^n\}$) is generated by the vector fields
\beq X^{(i)} \ = \ \frac{\pa}{\pa x^i}  \ , \ \ \ Y^{(i)} \ = \ e^{- \b t} \ \( \frac{\pa}{\pa x^i} \ - \ \b \, \frac{\pa}{\pa v^i} \)  \ . \eeq

Now, instead than going through the long and detailed steps followed in the case of linear force, we look directly for rectifying variables, i.e. for variables $(z^i, y^i)$ such that
\beq X^{(i)} ( z^i ) \ = \ 0 \ , \ X^{(i)} (y^i ) \ = \ 1 \ ; \ \ \ \
Y^{(i)} ( z^i ) \ = \ 1 \ , \ Y^{(i)} (y^i ) \ = \ 0 \ . \eeq By a standard computation (with the method of characteristics) these are given by
\beq z^i \ = \ - \ \frac{e^{\b t}}{\b} \ v^i \ , \ \ \ y^i \ = \ x^i \ + \ \frac{v^i}{\b} \ . \eeq

We can then compute, by a standard application of Ito formula, the equations satisfied by $(z^i, y^i)$ if $(x^i,v^i)$ evolve according to \eqref{eq:FC}. These turn out to be
\begin{eqnarray}
d z^i &=& - \ \frac{e^{\b t}}{\b} \ \[  c \, dt \ + \ \mu \, dw \] \, \\
d y^i &=& \frac{1}{\b} \ \[ c \, dt \ + \ \mu \, d w \] \ . \end{eqnarray}
Both equations are immediately integrated, the solutions being expressed in terms of Ito integrals,
\begin{eqnarray}
z^i(t) &=& z^i (t_0) \ - \ \frac{c^i}{\b^2} \ \( e^{\b t} \ - \ e^{\b t_0} \) \ - \ \frac{\mu}{\b} \ \int_{t_0}^t e^{\b t} \ d w^i (t) \ , \\
y^i (t) &=& y^i (t_0) \ + \ \frac{c^i}{\b} \ (t - t_0) \ + \ \frac{\mu}{\b} \ \int_{t_0}^t d w^i (t) \ . \end{eqnarray}

%\newpage

\section{Conclusions}

In the first part of the paper, up to Sect.\ref{sec:integr_reduct}, we have recalled some basic facts about invariants and symmetries of stochastic equations, and their use.

We have then studied, from Sect.\ref{sec:Ito_OU} on, the Ornstein-Uhlenbeck process in an external field, paying attention to invariants and to determination of its symmetries. The specific form of the equations \eqref{eq:BM} under study implies the presence of ``ghost'' (as opposed to ``true'' or ``real'') symmetries, related to variables appearing in the general formalism but not in the equations; these have not been considered. Moreover, we are only interested in \emph{simple} Lie symmetries, in the sense -- and for the reasons -- discussed in Sect.\ref{sec:symmstoch}.

We have first discussed the problem in full generality, i.e. in dimension $n$ and for the possibly non-isotropic case. For invariants this led to general conclusions in the case of constant or linear force; see Lemmas \ref{sec:general}.1 and \ref{sec:general}.2. For symmetries, this led to reduced formulas for the possible functional form of coefficients in symmetry vector fields, see \eqref{eq:redxifin}  and \eqref{eq:redetafin}; and to reduced determining equations, see \eqref{eq:redxigen} and \eqref{eq:redetagen}. A complete result can be obtained in the case of constant force field, see Lemma \ref{sec:general}.3, but little can be said for a general system.

We have then passed to consider more specifically the isotropic case; see Remark 16 for the exact sense of ``isotropic'' in this context.

We have preliminarily discussed the one-dimensional case, giving full details of the computations. In this case it turns out that {nontrivial invariants exist only for a constant force field, see Lemma \ref{sec:1Dsymm}.1.} As for symmetries, for a nonlinear force field $F(x)$ there are no simple symmetries, while for $F(x)$ linear there are two (commuting) real simple Lie symmetries { and a W-symmetry; finally} for $F(x)$ constant (in which case we know a nontrivial invariant $\chi$ exist) we have an infinite dimensional Lie symmetry algebra {which includes a sub-algebra of simple symmetries} with the structure of a Lie module over $\mathcal{C}^\infty (\chi)$; the infinite dimensional Lie algebra has an (infinite dimensional) Abelian ideal, which is also a Lie submodule. These results for the one-dimensional case are summarized in Lemmas \ref{sec:1Dsymm}.2 through \ref{sec:1Dsymm}.4.

We have then considered the general $n$-dimensional isotropic case. The discussion and the required computations are conceptually similar to those needed in the one-dimensional case; we have thus been less detailed in reporting our computations. On the other hand, the concept of nonlinear and linear forces are not sufficient to classify the different possibilities, unless we restrict ourselves to considering regular and second order regular force fields, as defined in Sect.\ref{sec:regular} {(see Definitions 1 and 2 therein)}. We found again that a (second order regular) nonlinear force field $\Fb (\xb)$ leads to no {regular} simple symmetries, {albeit it may have W-symmetries}  (Lemma \ref{sec:n_dim}.1), while for (regular) linear $\Fb (\xb )$ we have an Abelian symmetry algebra {of regular simple symmetries} of dimension $2 n$ (Lemma \ref{sec:n_dim}.2). {Considering also W-symmetries we still have an Abelian algebra (Lemma \ref{sec:n_dim}.3). Finally,} in the case of constant force fields, we have an infinite dimensional Lie symmetry algebra with the structure of a Lie module of dimension $2 n$ over $\mathcal{C}^\infty (\chi)$; again the infinite dimensional Lie algebra has an (infinite dimensional) Abelian ideal, which is also a Lie submodule (Lemma \ref{sec:n_dim}.4).

{As emerges from this summary of our discussion, the case of special non-linear -- and possibly non-regular -- problems possessing symmetries needs to be studied separately. It is maybe worth mentioning explicitly, in this context, that the notion of regular force field leaves out physically relevant cases: e.g. the Kepler problem is not regular.}

The results obtained here are preliminary to integration of the Ornstein-Uhlenbeck process for particles in an (isotropic) external field. This is based on the -- by now well established -- symmetry theory of stochastic differential equations, and in particular on \emph{Kozlov theory}  \gcite{Koz1,Koz2,Koz3}. Obviously this integration can take place only if symmetries are present, e.g. in the case where the force field is linear or constant. In Sect.\ref{sec:integration} we have shown indeed that in these cases, passing to \emph{symmetry-adapted variables}, integration is straightforward.

%\newpage

\label{lastpage}

\end{document}